\documentclass[12pt,preprint]{aastex}
\shorttitle{Wolf 630}
\shortauthors{Bubar et al.}

\begin{document}

\title{Spectroscopic Abundances and Membership in the Wolf 630 Moving Group}
\author{Eric J. Bubar}
\affil{Department of Physics and Astronomy, University of Rochester,
    P.O. Box 270171, Rochester, NY 14627-0171}
\email{ebubar@gmail.com}
\author{Jeremy R. King }
\affil{Department of Physics and Astronomy, Clemson University,
    Clemson, SC 29630-0978}
\email{jking2@ces.clemson.edu}

\begin{abstract}
The concept of kinematic assemblages evolving from dispersed stellar clusters has
remained contentious since Eggen's initial formulation of moving groups in the
1960's.  With high quality parallaxes from the Hipparcos
space astrometry mission, distance measurements for thousands of
nearby, seemingly isolated stars are currently available.  With these distances, a
high resolution spectroscopic abundance analysis can be brought to bear on the
alleged members of these moving groups.  If a structure is a relic of
an open cluster, the members can be expected to be monolithic in age and abundance
inasmuch as homogeneity is observed in young open clusters.  In this work we have 
examined 34 putative members of the proposed Wolf 630 moving group using high resolution 
stellar spectroscopy.  
The stars of the sample have been chemically tagged
to determine abundance homogeneity and confirm the
existence of a homogeneous subsample of 19 stars.  Fitting the homogeneous
subsample with Yale-Yonsei isochrones yields a single evolutionary
sequence of $\sim$2.7 $\pm$ 0.5 Gyr.  It is concluded that this 19 star
subsample of the Wolf 630 moving group sample of 34 stars could represent a dispersed 
cluster with an $<$[Fe/H]$>$=-0.01 $\pm$ 0.02
and an age of 2.7 $\pm$ 0.5 Gyr.  In addition, chemical abundances of Na and Al in giants
are examined for indications of enhancements as observed in field giants of old open
clusters, overexcitation/ionization effects are explored
in the cooler dwarfs of the sample and oxygen is derived from the infrared triplet and the 
forbidden line at $\lambda$6300 \AA. 
\end{abstract}


\section{INTRODUCTION}

A major goal of modern astronomy is to piece together 
the dynamic and chemical evolution of the Galactic disk.  To this end, one of the principle 
approaches for probing the disk has been to study open 
clusters.  Clusters are valuable astrophysical tools as they share 
common distances, common ages and common initial chemical abundances.
With the disk richly populated by both field stars and open clusters, and considering that clusters
are relatively well studied, the logical step in piecing together a more complete picture
of the chemical and dynamical history of the disk is to study field stars.

In recent years, the advent of large surveys such as \emph{HIPPARCOS} \citep{1997A&A...323L..49P} 
has yielded precise parallaxes for thousands of nearby field stars, and in doing
so, provided the necessary tools for investigating the field.  In particular, studies of 
the velocity distributions of disk field stars in the solar neighborhood have identified 
stellar overdensities in kinematic phase space (\citet{1999MNRAS.308..731S}).  The potential 
application of these velocity structures, commonly referred to as moving groups, was 
first identified by \citet{1958MNRAS.118...65E} who considered these assemblages to be 
relic structures of dissolved open clusters.
In this paradigm, a moving group is essentially a spatially unassociated open cluster;
therefore it should possess some of the same characteristics that make 
open clusters such valuable astrophysical tools (common ages and
common initial chemical abundances) and similar techniques that are useful for studying open clusters
could be applied.  

Relatively little work has been done to explore the 
reality of smaller moving groups (kinematic assemblages of $\sim$ 100 stars) 
as dissolved open clusters and their use in chemically 
tagging the galactic disk, with two notable exceptions: the Ursa Major Group and the HR1614
Moving Group.  \citet{1993AJ....105..226S} examined the Ursa Major moving group
and utilized age information inferred from chromospheric emission to constrain group membership
in UMa.  While this study did not utilize chemical tagging to constrain group membership, it did illustrate the
viability of moving groups as dissolved populations of open clusters.  \citet{2003AJ....125.1980K} and
\citet{2005PASP..117..911K}
revisited the membership of the UMa group, using new and extant abundances.  They used the 
results to constrain membership in the UMa group, showed the members to be chemically homogeneous, 
and noticed overexcitation/overionization effects in the cooler field star members of the group, 
similar to those observed in young ($<$ 500 Myr) cool open cluster dwarfs (\citet{2003AJ....125.2085S},
\citet{2004ApJ...602L.117S}).  The first in depth application of chemical tagging to constrain 
moving group membership was by \citet{2007AJ....133..694D}, who derived 
abundances for various elements for the HR 1614 moving group.  They found that for their 18 star 
sample, 14 stars were metal-rich ([Fe/H] $\geq$ 0.25 dex with $\sigma$=0.03) leading to 
the conclusion that the HR 1614 moving group, with its distinct kinematics and distinctly super-solar
chemical abundances, was a remnant of a dissolved open cluster.

In the field of moving group populations, the classical Wolf 630 moving group is an intriguing target.	
The first identification of the Wolf 630 moving group was made by
\citet{1965Obs....85..191E} who noted that several K and M dwarfs and giants 
in the solar neighborhood appeared to have similar space motions to that of the multiple star
system Wolf 630 ((U,V,W)=(23, -33, 18) kms$^{-1}$).  These kinematics, distinctive of
membership in an old disk population, placed the stars in a relatively sparsely populated
region of kinematic phase space \citep{1969PASP...81..553E}.  Eggen also noted that the 
color magnitude diagram for the K and M dwarfs and giants with kinematics similar to those of Wolf
630 appeared to trace an evolutionary sequence similar to the old ($\sim$ 5 Gyr; \citet{1999AJ....117..330J})
M67 open cluster .  Although his sources are not completely transparent, at least
some (17 of 54 stars) of the distances in his study were determined from trigonometric parallaxes , 
with the remainder coming from ``luminosity estimates of many kinds''.
As a rudimentary form of chemical tagging, \citet{1970PASP...82...99E} estimated metallicities 
of 23 Wolf 630 group members through \emph{uvby-$\beta$} photometry.  Variations in the 
$\delta$[m$_1$] index were found to be comparable 
to the Hyades, Praesepe, and the Coma Berenices clusters, implying chemical homogeneity.  

\citet{1979LIACo..22..355T} studied the chemical 
composition of five field giant stars that were alleged members 
of Wolf 630 using high dispersion coud\'{e} spectra described in \citet{1979LIACo..22..355T}.
Employing a curve of growth approach and measured equivalent widths, they  
found that three stars appeared to be chemically homogeneous with an overall metallicity for Wolf 630 
of [Fe/H]$\sim$+0.23.  However, it must be noted that their abundances were not measured with
respect to the Sun, but are instead quoted with respect to a standard star of presumed
solar metallicity (HD 197989), which has since been determined to be a K0III.  While they derived a metallicity 
of 0.00 for their reference star, literature determinations suggest a value of -0.24.  
This would lower the average metallicity for the group to [Fe/H]$\sim$ -0.02.

\citet{1983MNRAS.204..841M} revisited the membership of the Wolf 630 moving group by
recreating the approach presumably
utilized by \citet{1965Obs....85..191E} to find his original Wolf sample.  In summary, they calculate the parallax
that yields a V velocity for each group candidate equal to the assumed group velocity of V=-32.8 $\pm$ 1.3 kms$^{-1}$.
The final absolute magnitudes they report assume these parallaxes.  Typical uncertainties in their 
absolute magnitudes appear to be between 0.2-0.4 magnitudes, larger than magnitude uncertainties obtainable
with precise parallax information currently available from Hipparcos.  The color-magnitude diagram assuming these
M$_{V}$ values was compared to the scatter of apparent members
with the observed scatter in the old open cluster M67.  They concluded 
that either (1) the intrinsic scatter in the Wolf 630 moving group color-magnitude diagram was greater
than that of M67, or (2) the errors in radial velocities and/or proper motions they utilized must have been
underestimated by a factor of 2.4 or (3) many of the stars in their sample were, in fact, non-members.  

\citet{1994AAS...185.4516T} examined metallicities from ``published values of [Fe/H] from diverse papers'' 
of 40 members of the Wolf 630 group.  His sample contains 26 \% of Eggen's original objects 
\citep{1969PASP...81..553E}.  He concluded that metallicity dispersions within his sample were too 
great for meaningful conclusions about the existence or non-existence of a genuine, chemically distinct 
Wolf 630 moving group.  This suggests the need to obtain high quality [Fe/H] determinations with 
minimal uncertainties in testing for chemical uniqueness in a putative Wolf 630 sample.

The analysis of solar neighborhood Hipparcos data by \citet{1999MNRAS.308..731S}
indicates a kinematic rediscovery of the Wolf 630 group.  Their figure 10, showing the \emph{UV} velocity distribution 
for 3561 late type dwarfs in the solar neighborhood presents a clear overdensity of stars near the 
position of Wolf 630.  Furthermore, this structure appears to be distinctly separated from any other 
known moving groups or stellar streams.  This provides compelling evidence that Wolf 630 is a real 
kinematic structure.  The question to be asked is if this kinematic structure is composed of stars
with a common origin?

Despite the distinctive kinematics exhibited by the Wolf 630 moving group when examined with updated
Hipparcos parallaxes, it has not been specifically targeted in an abundance study which makes use of the modern
astrometric and spectroscopic data.  This is remedied in this paper, where accurate parallaxes and photometry from 
the updated \emph{HIPPARCOS}  
data reduction \citep{2007A&A...474..653V} coupled with high precision radial velocities from CORAVEL 
(\citet{2004A&A...418..989N} and references therein) allow for developing a Wolf 630 sample 
with internally consistent distances and absolute magnitudes, thereby removing the uncertainties faced by 
\citet{1983MNRAS.204..841M}.  Furthermore, our uniform high resolution spectroscopic study of 
Wolf 630 moving group candidate members provides a single, consistent 
set of metallicites with low internal uncertainty to test chemical homogeneity in the 
group, removing the largest source of uncertainty from \citet{1994AAS...185.4516T}.  
    
 \section{DATA, OBSERVATIONS AND ANALYSIS}
\subsection{Literature Data}

The 34 stars in this sample, listed in Table 1, were previously identified as members of the Wolf 630 
group (\citet{1969PASP...81..553E}, \citet{1983MNRAS.204..841M}) according
to their \emph{UVW} kinematics.  In this study, we use updated parallaxes and proper motions from the latest 
reduction of Hipparcos data \citep{2007A&A...474..653V}.  Precision radial 
velocities were taken from the compilation of \citet{2004A&A...418..989N}.  Visible band photometry 
(\emph{B}, \emph{V}, \emph{B$_{Tycho}$}, \emph{V$_{Tycho}$}) was taken from the 
{\it HIPPARCOS\/} catalogue \citep{1997A&A...323L..49P}.  Near infrared \emph{J}, \emph{H} and 
\emph{K} photometry was taken from the 2MASS Catalog \citep{2003tmc..book.....C}.  

\subsection{Kinematics}

We determined galactic \emph{UVW} kinematics from the proper motions, parallaxes and
radial velocities using a modified version of the code of \citet{1987AJ.....93..864J}.
Here, the U velocity is positive towards the Galactic center, the V velocity
is positive in the direction of Galactic rotation and the W velocity is
positive in the direction of the North Galactic Pole (NGP). 
The relevant parameters for determination of these kinematics are presented 
in Table \ref{uvw}.  

\subsection{Spectroscopic Observations and Reductions}

Spectroscopy of the sample was obtained in March 2007 
and November 2008 with the KPNO 4 meter Mayall telescope, the echelle spectrograph 
with grating 58.5-63 and a T2KB 2048X2048 CCD detector.  The slit width of $\sim$ 1 
arcsec yielded a resolution of R $\sim$ 40,000 with a typical S/N of 200 per summed pixel.  The spectra 
have incomplete wavelength coverage extending from approximately 5800 {\AA}  to 7800 {\AA}.  
The spectra have been reduced using standard routines in the {\sf echelle} package of 
IRAF{\footnote{IRAF is distributed by the National Optical Astronomy Observatories, 
which are operated by the Association of Universities for Research in Astronomy, Inc., 
under cooperative agreement with the National Science Foundation.}.  These include 
bias correction, flat-fielding, scattered light correction, order extraction, and 
wavelength calibration.  Sample spectra are presented in Figure \ref{spectra}.

\subsection{Line Selection}

Spectroscopic physical parameters are typically determined by enforcing balance constraints
on abundances derived from lines of Fe, which has a plethora of suitable neutral (\ion{Fe}{1}) 
and ionized (\ion{Fe}{1}I) features in the optical.  We compiled low excitation potential 
($\chi <$ 6.00 eV) \ion{Fe}{1} and \ion{Fe}{2} lines from \citet{1990A&AS...82..179T}, 
the VIENNA Atomic Line Database (\citet{1995A&AS..112..525P}, \citet{1997BaltA...6..244R}, 
\citet{1999POBeo..65..223K}, \citet{2000BaltA...9..590K}), \citet{2004ApJ...603..697Y}, 
\citet{2006AJ....131.1057S} and \citet{2007AJ....133..694D}.  
Lines that were not apparent in a high-resolution solar spectrum \citep{2005MSAIS...8..189K} 
were removed from the linelist.  
In order to guarantee that these lines were unaffected by blending effects, especially those
arising in cool stars that might not be noticeable in the solar spectrum, 
the 2002 version of the MOOG spectral analysis program \citep{1973PhDT.......180S} was used to 
compute synthetic spectra in 1 {\AA}  blocks surrounding all Fe features, using VALD linelists.  
If a line had closely neighboring features with MOOG-based relative strength parameters within an order of magnitude 
it was removed from consideration.  In this manner, a final list of 145 \ion{Fe}{1} lines and 11 \ion{Fe}{2} lines
was formed.  These linelists are presented in Table \ref{eqw}.  The equivalent
widths listed are for measurements in a high resolution solar spectrum.  

Linelists for other elements of interest have also been compiled from multiple 
sources (\citet{1990A&AS...82..179T}, \citet{1998AJ....115..666K}, \citet{2006AJ....131..455D}).  
These elements include Li, Na, Al, Ba, a selection of $\alpha$ elements (O, Mg, Si, Ca, Ti I and Ti II)
and a selection of Fe peak elements (Cr, Mn and Ni).  The lines are also given in Table \ref{eqw}.
Equivalent widths are again for
measurements in the high resolution solar spectrum.  The equivalent widths 
that were measurable for each individual star are given in Table \ref{alleqw}, with
corresponding abundances derived from each equivalent width.

\subsection{Equivalent Widths}

Equivalent widths for the lines of interest were measured in each star and in a high resolution solar spectrum
using the spectral analysis tool SPECTRE \citep{1987BAAS...19.1129F}.   
Final abundances were obtained from equivalent widths through use of the MOOG LTE spectral
analysis tool \citep{1973PhDT.......180S} with an input
Kurucz model atmosphere characterized by the four fundamental physical parameters: temperature, surface gravity, 
microturbulent velocity ($\zeta$) and metallicity.  Unless noted otherwise all abundances are differential with respect to 
the Sun and are presented in the standard bracket notation 
([X/H]=log($\frac{N(X)}{N(H)}$)$_* - $ log($\frac{(N(X)}{N(H)}$)$_\odot$ where logN(H)$\equiv$ 12).

\subsection{Initial Parameters: Photometric  }

The color-$T_{\rm eff}$-[Fe/H] calibrations of \citet{2005ApJ...626..465R} were used to
determine photometric temperatures from Johnson $B-V$, Tycho $B_T-V_T$, Johnson/2MASS $V-J_2$,
$V-H_2$ and $V-K_2$.  The color indices for 8 stars were
outside of the calibrated ranges; consequently
photometric temperatures were not derived.  Uncertainties in the photometric
temperatures were conservatively taken as the standard deviation of the temperatures
derived from each of the respective colors.
With the availability of high quality Hipparcos Parallaxes, physical
surface gravities were calculated from:

\begin{centering}
$log\frac{g}{g_{\odot}} = log\frac{M}{M_{\odot}}+4log\frac{T_{eff}}{T_{eff,\odot}}+0.4V_o+0.4B.C.+2log\pi+0.12$\\
\end{centering}

\noindent where M is the mass in solar masses, estimated from Yale-Yonsei isochrones \citep{2004ApJS..155..667D}
of solar metallicity, bolometric corrections are from \citet{2005oasp.book.....G}, and 
$\pi$ is the parallax.  Initial microturbulent velocities were found from the calibrations of \citet{2004A&A...420..183A}.
These photometric parameters provided the initial guesses for physical parameters when deriving the final
spectroscopic values.  Additionally, the photometric calibrations provided reasonable estimates to compare to
spectroscopically derived results.  Using the updated calibrations of \citet{2010A&A...512A..54C} does not change the 
results described herein.

\subsection{Spectroscopic Parameters}

The refined Fe linelists discussed above acted as target lists for each 
of the stars in the sample.  The typical star contained $\sim$80 of the 145 good \ion{Fe}{1} lines 
that were measurable in the solar spectrum.  
Several of the stars showed correlations between \ion{Fe}{1} excitation potential and reduced 
equivalent width.  If ignored, such correlations can be imposed onto the temperatures and microturbulent 
velocities, resulting
in non-unique solutions of physical parameters.  Consequently, two 
linelists for the \ion{Fe}{1} lines were formed for each star; a correlated and an uncorrelated sample.
Final basic physical parameters for the sample were derived using a modification to the standard techniques
of Fe excitation/ionization/line strength balance.  In all the approaches
described below, a differential analysis was used where the same lines were measured in a solar
spectrum and in the stellar spectra.  Final abundances were then determined by subtracting the solar
abundance from the stellar abundance in a line by line fashion.

The first technique utilized the uncorrelated line sample and proceeded as follows:
temperatures of input model atmospheres were adjusted to remove
any correlation in solar-normalized abundances with respect to excitation potential; $\zeta$ is adjusted to remove
any correlation with line strength and log $g$ is adjusted until the mean abundance from \ion{Fe}{1}I lines matches
the abundance from \ion{Fe}{2} lines. This approach required
simultaneously adjusting temperatures, surface gravities, metallicities and microturbulent velocities to 
converge to a common solution.  Use of the uncorrelated line sample, as described above, is necessary 
to ensure a unique solution.  This approach will be referred to as the ``classical'' approach.

The second approach used the correlated line sample and the Hipparcos-based physical surface gravities.  
The \ion{Fe}{2} abundances are primarily set by this gravity.  
The temperature was adjusted to force the mean abundance from \ion{Fe}{1} lines to match that from \ion{Fe}{1}I lines.  
The microturbulent velocity was adjusted until the abundance from \ion{Fe}{1} lines had no dependence on reduced equivalent 
width.  The advantage of this approach is that it does not require simultaneous solutions requiring excitation balance 
and equivalent width balance, allowing use of a full correlated line sample.  This approach will be 
referred to as the ``physical surface gravity'' approach.

When comparing results from the classical and physical surface gravity approaches it was apparent that the 
microturbulent velocities were nearly identical ($\delta \zeta \approx \pm 0.04$ km s$^{-1}$. 
Thus our final spectroscopic parameters were determined as follows.  
The microturbulent velocities from the ``classical'' approach and the 
``physical surface gravity'' approach were averaged to yield a final value.  The correlated line sample 
was used to determine the temperature and surface gravity using excitation/ionization balance.  
For the remainder of the work, the results from this approach were used for the physical parameters of these 30 
stars.  The remaining 4 stars would not converge to an acceptable solution and the following alternative approach 
was developed.

The coolest stars in the sample (HIP105341-dwarf, HIP114155-giant and HIP5027-dwarf) had an insufficient 
number of well-measured \ion{Fe}{2} lines for accurately determining the surface gravity 
spectroscopically.   Consequently, Hipparcos-based physical surface gravities were used to set the gravity, and 
the temperature and microturbulence were iterated to eliminate correlations in [\ion{Fe}{1}/H] versus excitation potential
and versus the reduced equivalent width.  This is the ``physical surface gravity'' approach.  

Finally, one of the stars in the sample (HIP 5027) had a microturbulence correlation which could not be removed 
without utilizing unreasonable surface gravities.  For this star, the surface gravity was set based on Yale-Yonsei 
isochrones \citep{2004ApJS..155..667D}.  The microturbulent velocity was set to zero and the temperature 
was determined from excitation balance. 

The final basic physical parameters 
(T$_{Spec}$, log $g$, microturbulent velocity ($\xi$) and [Fe/H]) are
presented in Table \ref{basic} and final abundances are summarized in Table \ref{abundances}.
For the interested reader, we also provide plots of all abundances ([X/H]) versus [Fe/H] in an appendix.

\subsection{Lithium}

Abundances have been derived for lithium using spectral synthesis.  We
use the {\sf synth} driver of MOOG to synthesize a spectrum of the lithium line
at $\lambda$=6707.79 {\AA} with an updated version of the linelist from \citet{1997AJ....113.1871K}. 
Appropriate smoothing factors were determined by measuring
clean, weak lines in the lithium region.  The lithium abundance was varied until a best fit is obtained 
from visual inspection.  A sample synthesis is presented in Figure \ref{lithium_synthesis}.

Uncertainties in lithium abundances have
been determined by examining the change in Li abundance in syntheses with arbitrary changes in
physical parameters of $\Delta$T=150 K, $\Delta$log\emph{g}=0.12 cm s$^{-2}$
and $\Delta$$\xi$=0.60 km s$^{-1}$, and adding the resultant abundance differences
in quadrature.  

\subsection{Oxygen}

Oxygen abundances for many stars have been derived from the near-IR $\lambda$7771 
equivalent widths. 
Abundances derived from the triplet are known to be enhanced by NLTE effects;
therefore appropriate corrections have been applied following 
\citet{2003A&A...402..343T}.  

For the giants and subgiants, oxygen abundances have also been derived from the
forbidden line at $\lambda$6300.34 \AA.  While this line is found to be free
from NLTE effects (\citet{2003A&A...402..343T}), care must be taken as the line is
blended with a nearby Ni feature at $\lambda$6300.31 \AA.  This blend is treated
using the {\sf blends} driver of MOOG, following \citet{2006AJ....131.1057S}.  
The  Ni abundance utilized to account for blending is the mean value derived from the
EWs of \ion{Ni}{1} lines in our sample.

A possible CN feature at 6300.265 {\AA} and two at 6300.482 {\AA} with log(gf)
values of -2.70, -2.24 and -2.17 are claimed by \citet{1963rspx.book.....D}.  
In order to explore these blends, multiple syntheses of the $\lambda$6300 {\AA} region 
were performed using high resolution spectral
atlases of the Sun \citep{2005MSAIS...8..189K} and the K giant, Arcturus \citep{2000vnia.book.....H}.  
The CN features, if real, were found to be unimportant in
the solar spectrum.  Large variations in carbon abundances ($>$ 0.50 dex) appear to have
little impact on the overall spectrum.  For warm dwarfs, the syntheses confirm
that the Ni features are the dominant blends affecting O determination.  

The situation appears to be dramatically different for cooler giants.  In the high resolution 
spectral atlas for Arcturus, the CN blend, if real, appears to dominate over the Nickel blend.
Oxygen syntheses were performed in order to calibrate the gf values of the CN molecules in the
linelist to match the spectrum of Arcturus, but results were inconclusive.  In particular, appropriate
smoothing factors were difficult to determine as the ultra high 
resolution of the Arcturus atlas makes Gaussian smoothing by an instrumental profile inappropriate.  
In an attempt to accurately reflect the smoothing in the
spectral atlas, broadening was done using a convolution of a macroturbulent broadening of 5.21 $\pm$ 0.2 kms$^{-1}$
\citep{1981ApJ...245..992G} and a rotational broadening characterized by vsin(i)=2.4 $\pm$ 0.4 kms$^{-1}$ 
\citep{1981ApJ...245..992G} with a limb darkening coefficient of 0.9 (from \citet{2005oasp.book.....G}).
With this smoothing, the spectrum for Arcturus in the forbidden oxygen region was fit by increasing the CN
features gf values by $\sim$0.40 dex, while assuming a [C/Fe]=-0.06 as found by \citet{2002AJ....124.3241S}.  
In attempting to apply this calibrated linelist to synthesize the forbidden line region for
one of the giants in our sample (HIP17792; chosen because its physical parameters were
similar to those of Arcturus) no reasonable abundance of carbon yielded a satisfactory fit.  This may suggest
that the gf values in the linelist need to be more well constrained.  In light of the ambiguous
results, it is concluded that an accurate determination 
of the carbon abundance is essential for proper treatment of any CN blending feature that may exist.  
We suggest that a spectroscopic analysis of cool giants with appropriate wavelength coverage to allow 
measurement of a precise carbon abundance would allow for calibration of the forbidden oxygen linelist, which
would be a project of not insignificant interest. 
Unfortunately the wavelength coverage of the observed spectra does not include any appropriate carbon features to
allow definitive conclusions as to the reality of the CN blending features found in \citet{1963rspx.book.....D}.
In light of the unresolved nature of this CN blending, abundances reported herein do not include it.  
Further justification for ignoring the CN blending is discussed in the results.

\subsubsection{Uncertainty Estimates}

The uncertainties in experimental and theoretical log(gf) values (likely at least
0.1 dex) can be a significant source of error; however, by performing a line-by-line differential analysis with respect
to the Sun, uncertainties due to transition 
probabilities are eliminated to first order.

Here, then, it is the uncertainty in physical parameters that underlie the uncertainties in the abundances.  
Errors in the temperature were
determined by adjusting the temperature solution until the correlation between [Fe/H] and excitation 
potential (excitation balance) reached a 1-$\sigma$ linear correlation coefficient for the given number of lines.  
The uncertainty in microturbulent velocity was determined in the same manner, by adjusting the microturbulence 
until the linear correlation coefficient for [Fe/H] versus equivalent width (equivalent width balance) 
resulted in a 1-$\sigma$ deviation.  For HIP 5027, which would not converge to a unique solution for microturbulence, 
an uncertainty in microturbulence of 0.20 kms$^{-1}$ was adopted.

For the cases where the physical 
surface gravity was utilized, the uncertainty was estimated by propagating the 
uncertainties in the temperature, mass, apparent magnitude, parallax and bolometric corrections.
The uncertainties in the spectroscopically determined surface gravities required a deeper treatment.  
Since gravity is calculated by eliminating the difference in iron abundance derived from [\ion{Fe}{1}/H] and [\ion{Fe}{2}/H],
the uncertainty in surface gravity is related to the quadratic sum of the the uncertainties in [\ion{Fe}{1}/H] and [\ion{Fe}{1}I/H].  
These abundances, in turn, have sensitivities that depend on the basic physical 
parameters.  Proper uncertainty calculations, therefore, require an iterative procedure.
The errors in [Fe I/H] and [Fe II/H] are a 
combination of the measurement uncertainties and the uncertainties in the 
physical parameters.  The line measurement uncertainties in Fe I and Fe II were estimated as the standard
deviation of the abundances from all Fe I and Fe II lines, respectively.  Abundance sensitivities for 
arbitrary changes in temperature ($\pm$ 150 K), surface gravity ($\pm$ 0.12 dex) and microturbulence 
($\pm$ 0.60 kms$^{-1}$) were determined by 
adjusting each parameter individually and recording the resultant difference in abundance.  To determine
abundance uncertainties the abundance differences must be properly normalized by the respective parameter's uncertainty.
For example, in HIP3455 the total temperature uncertainty was
found to be 35 K.  The final abundance uncertainty introduced by the arbitrary temperature change would, therefore, be
equal to the difference in abundance multiplied by $\frac{35 K}{150 K}$, where 35 K is the temperature uncertainty and
150 K is the arbitrary temperature change introduced to determine the temperature sensitivity.
For the first calculation the uncertainties in temperature and microturbulent velocity were determined as 
above and the uncertainty in surface gravity was unknown; consequently its contribution to abundance
uncertainty was initially ignored.  Adding the measurement errors in [Fe I/H] and [Fe II/H] in quadrature
with the physical parameter abundance uncertainties from temperature and microturbulence yields
a first estimate for the uncertainty in the surface gravity.  This gravity uncertainty can then be added
in quadrature to the line measurement uncertainty, the temperature uncertainty and the microturbulent uncertainty
to yield a final uncertainty for the surface gravity.  The surface gravity in the model atmosphere was adjusted
until the difference in abundance between [Fe I/H] and [Fe II/H] was equal to their quadrature added
uncertainties.  The difference between this gravity and the spectroscopically derived gravity provides the final
uncertainty in surface gravity.

Uncertainties in abundances were found by introducing arbitrary changes in T, microturbulence 
and surface gravity ($\Delta$ T=150 K, $\Delta$ $\xi$=0.60 kms$^{-1}$, and $\Delta$log $g$=0.12 cm s$^{-2}$),
normalized by the respective parameter uncertainties.  
The uncertainties introduced by each of these parameter changes was added in quadrature to obtain 
total parameter-based uncertainties.  Measurement uncertainties were taken as the uncertainty
in the weighted mean for all lines of a given element.  For elements with only a single line available, the 
standard deviation of all \ion{Fe}{1} abundances was utilized as an estimate of the line measurement uncertainty.  
The final uncertainties in the abundances were determined by adding the parameter-based abundance uncertainties 
with the measurement uncertainties in quadrature.

A sample table of the normalized parameter changes and their final resultant 
[\ion{Fe}{1}/H] errors on a given star is presented in Table \ref{errors}.  

\subsection{Physical Parameter Comparisons:} 
\subsubsection{Temperatures: Spectroscopic Versus Photometric}

The temperatures for the stars in the sample were determined from photometric calibrations as
well as through spectroscopic excitation balance.  In Figure \ref{param_diff}
the spectroscopic temperature is plotted versus the photometric temperature.  The line represents perfect agreement
between the two temperatures.  It can clearly be seen that
the temperatures from the two techniques are equivalent within their respective uncertainties.  
There is a slight indication that spectroscopic 
temperatures may be systematically higher, with 66 \% of the stars lying above the line, however the effects
on the abundance analysis are negligible and do not change any conclusions.     							

\subsubsection{Surface Gravity: Spectroscopic Versus Physical}

The surface gravity was determined from Hipparcos data (i.e. physical surface gravities) and 
spectroscopically via ionization balance.  In 
Figure \ref{param_diff}, the spectroscopic surface gravity is plotted versus 
the physical surface gravity.  The line shows the trend for the gravities being equal.  
Within their respective 
uncertainties, the surface gravities are equal.

\section{RESULTS AND DISCUSSION}

The primary goal of the paper is to determine if the kinematically defined
Wolf 630 Moving Group represents a stellar population 
of a single age and chemical composition.  The sample
stars have been plotted in the HR diagram (Figure \ref{HR_spec_final}) to determine if they are
coincident with a single evolutionary sequence.  The sequence
traced by the majority of stars coincides with a Yale-Yonsei 
isochrone \citep{2004ApJS..155..667D} of 2.7 $\pm$ 0.5 Gyr with an assumed solar metallicity.  
In attempting to qualitatively use ages as a constraint for establishing
membership in a distinct evolutionary sequence, it will be assumed that the isochrone
which fits the majority of the sample provides a reasonable estimate of the age range of 
a dominant coeval group, if it indeed exists.

The abundance results are presented in Table \ref{abundances} and as plots of [X/H] versus temperature (Appendix).  
Lithium and oxygen abundances were also derived, but they are presented and discussed
separately as the approach utilized for these abundance results involved synthesis (Li) or use
of the MOOG \emph{blends} driver (O).  In order to visually present the abundance results, the metallicity distribution
of the entire sample is presented in the form of a ``smoothed histogram'' in Figure \ref{fe_smooth}.
This distribution has been generated by characterizing  each star with a gaussian 
centered on its mean [Fe/H] with standard deviation equal to the [Fe/H] uncertainty.  
The distributions are summed to yield a final smoothed histogram and have been renormalized to a unit
area.  In this manner, the distributions include uncertainties 
in abundances, making them useful for a visual examination of the complete sample to discern
if any stars yield abundances that deviate from the sample as a whole.  The distribution
is clearly not unimodal or symmetric.  It is dominated by a near-solar metallicity peak and two 
smaller peaks at [Fe/H]$\sim$-0.50 and [Fe/H]$\sim$+0.30.  It is clear that our Wolf 630 
moving group sample is not characterized by a single chemical composition.

\subsection{Approach to Chemically Tagging}

While our entire sample cannot be characterized
by a single chemical abundance, we can investigate whether there is a dominant subsample having
common abundances and age.  
This is done by eliminating stars that are clearly outliers, using arguments
based on extreme abundances, evolutionary state (inferred from HR diagram positions,
lithium abundance, chromospheric acitivities and surface gravities) or a combination thereof.
These members will be classified as ``unlikely'' members of a dominant homogeneous group.  In this
way we can, for example, establish a subsample that is characterized by a dominant [Fe/H], if it exists.
Stars with such an [Fe/H] will be classified as
either ``possible'' or ``likely'' members of a chemically homogeneous, isochronal population
having common kinematics.  The final
distinctions between ``possible'' and ``likely'' will be made based on evolutionary status
and additional abundance information inferred from lithium, alpha elements and 
iron peak elements.  Particular interest is paid to the iron abundance, [Fe/H], as it is
considered the most well determined abundance, primarily 
due to the quality and size of the Fe line sample. 

The quantitative constraint adopted for determining chemical homogeneity
was to require that a star's abundance, within its
uncertainty, rest within a metallicity band centered on the weighted mean abundance of
stars in the sample.  The half-width of this band was conservatively taken to be 
3 times the uncertainty in the weighted mean.  This approach was followed in an iterative fashion
where whenever a star was determined to be an ``unlikely'' member of a dominant chemical group
it was removed from the sample and a new weighted mean and band size was found.  In this
manner, a common abundance for the sample was converged to for each element (except Lithium and Oxygen).
Examples of the band plots for [Fe/H] versus T$_{\rm eff}$ is given in Figure \ref{fe_abtemp}, 
where [Fe/H] is plotted versus temperature.  The solid line gives the weighted mean [Fe/H] while the 
dotted lines give the 3-$\sigma$ uncertainties in this mean, i.e. the abundance band.  

This visual analysis from examining the abundance distributions served as a guide for identifying the clearly
unlikely members.  Abundance information
alone was used to constrain giant star membership in a dominant chemical group, as robust discriminants of age
are unavailable.  Many of the dwarfs lay above the main sequence, leading
to the question of if they might be pre-or-post main sequence objects.    
Consequently a diagnostic was needed to constrain evolutionary status for these dwarf and subgiant stars. 
The full analysis, therefore, examined each star individually, utilizing abundances and 
information on evolutionary status (inferred from chromospheric activities, isochrone ages and surface gravities)
to classify each star in its appropriate category (unlikely, possible or likely).      

Figure \ref{lithium} shows the absolute lithium abundance
versus effective temperature for the litle-evolved stars in our sample and for a sample of dwarf stars in the 
Pleiades, Hyades, NGC752 and M67.  The lithium abundances of the sample stars are plotted with each cluster: 
filled hexagons are dwarfs, filled triangles are upper limits for dwarfs, open hexagons are subgiants (as inferred
from HR-diagram positions and apparently low levels of chromospheric activity) and open triangles are
upper limits for subgiants.  Accepted ages are given for each of the respective clusters, with the 
Pleiades trend being used as a baseline to indicate that a star is likely to be young (i.e.
if a star has a lithium abundance which rests in the Pleiades lithium abundance trend it is likely
a young star).

\subsection{Final Membership}

With the considerations above, the 34 stars in the sample have been classified as unlikely, possible and
likely members of a common chemical, temporal and kinematic assemblage.  There were a total
of 13 stars removed from group membership due to classification as unlikely members.  
If the remaining 21 stars classified as possible and likely are considered to represent a chemically
distinct group, then out of the original kinematically defined sample, $\sim$ 60\% remain members
of a kinematically and chemically related group with a common 2-3 Gyr age insofar as we can tell.  

The final evolutionary
sequence traced by the possible and likely members is presented in Figure \ref{HR_spec_final}, with
possible members plotted in red and likely members plotted in green.  The group is reasonably well traced 
by an evolutionary sequence of $\sim$ 2.7 Gyr (solid line) with lower and upper limits of 2.2 Gyr and 3.2 Gyr 
(dashed lines).   The dwarf members, HIP 41484, HIP 105341, HIP 14501 and HIP 43557,
have positions that place them slightly above the main sequence; however, based on lithium abundances, none of the
stars are believed to be pre-main sequence objects and surface gravities are all consistent with dwarf status.  
The giants HIP 3992, HIP 34440 and HIP3455 appear
to form a red giant clump.  The remaining members all lay on the best fit isochrone within their 
respective uncertainties.  Thus the possible and likely members we identify can be
characterized by a distinct evolutionary sequence of 2.7 $\pm$ 0.5 Gyrs.

The final UV kinematic phase space plot is presented in Figure \ref{uvw_final}, where possible members 
are again red and likely members are green.  For our initial full sample, the RMS U and V velocities
are 23.92 and 34.46 kms$^{-1}$, respectively.  In the final subsample of group members, U$_{RMS}$=25.21 kms$^{-1}$
and V$_{RMS}$=35.8 kms$^{-1}$, therefore the kinematic identity has not been 
significantly altered by the requirement of chemical and temporal coherence to establish group membership,
which points to the necessity to utilize criteria other than kinematics to robustely link members of
moving groups.

The weighted mean abundances of the final possible and likely members of a dominant chemical group
are presented in Table \ref{group}.  The quoted errors are the uncertainties in the weighted
mean.  In order to explore the homogeneity of our samples a reduced chi-squared statistic 
is presented for each element assuming a constant mean abundance.    
Performing this test for [Fe/H] for warm stars (T$\ge$ 5000 K) in the Hyades cluster sample data 
from \citet{2006AJ....131.1057S}, yields a $ \chi_{\nu}^{2}$ of 1.303.  
For a set of 7 Pleaides stars from \citet{2003AJ....125.2085S}, 
the reduced chi-squared in [Fe/H] is 1.818.  Note that the cool stars were removed 
from the calculation as they are believed to be impacted by overexcitation/ionization effects.
From these chi-squared values, we estimate the Hyades and Pleiades are chemically homogeneous
with a roughly 2-sigma significance.  With these open clusters assumed to be chemically homogeneous,
an approximate reduced chi-squared of $\le$ 2, therefore, provides a rough quantitative indication of homogeneity.
The $ \chi_{\nu}^{2}$
is presented for the full sample of 34 stars ($\chi_{\nu}^2all$), the final sample of 21 possible
and likely group members ($\chi_{\nu}^2group$) and the 11 likely members ($\chi_{\nu}^2likely$).  
First, the very large $\chi_{nu}^{2}$ for the full sample confirms that the initial kinematically defined
sample of alleged Wolf 630 members is clearly not chemically monolithic.  
The decrease in reduced chi-squared between the full sample and the chemically
distinct subsample demonstrates that chemically discrepant stars have been removed.  Even in the
likely subsample the $\chi_{\nu}^2$ values remain uncomfortably large for Na and Al.  
Discussion of these discrepancies is reserved for a later section.

Considering the reduced chi-squared for other homogeneous open cluster samples
is comparable to the reduced chi-squared for the possible and likely members of the
sample across multiple elements, the chosen sample is considered to represent a
chemically consistent group with a weighted average metallicity of 
[Fe/H]=-0.01 $\pm$ 0.02 (uncertainty in the weighted mean).
{\textbf Using precise chemical tagging of the 34 star sample of the Wolf moving group,
a single evolutionary sequence of 2.7 $\pm$ 0.5 Gyr and [Fe/H]=-0.01 $\pm$ 0.02
has been identified for a subsample of 19 stars.}

\subsection{Open Clusters and Moving Groups: Chemically Tagging the Disk}

We present additional results here that illustrate the application
of moving group field star members in exploring stellar and chemical evolution in the Galactic disk.

\subsubsection{Na and Al Abundances}

The abundances of Na and Al appear to be enhanced for some of the stars in the sample.
Similar enhancements have been observed in many open clusters.  Most recently,
an analysis of abundances in the Hyades cluster found abundance enhancements
in Na and Al of 0.2-0.5 dex in giant stars when compared with dwarfs (\citet{2009ApJ...701..837S})
in line with observations of giant stars in old open clusters (\citet{2005AJ....129.2725F},
\citet{2008AJ....135.2341J}).
These enhancements can be compared to those observed in the group members of this work.

Plots of [Na/Fe] (top panel) and [Al/Fe] (bottom panel) versus surface gravity are presented in Figure \ref{na_al}.
For the members of the group, the Na and Al enhancements are relatively modest, as seen
in a relatively slight upward shift in abundances between dwarfs and subgiants.  The giant abundances,
in general, can be brought into agreement with dwarf abundances with 
downward revisions of 0.1-0.2 dex, consistent with NLTE corrections found in field clump 
giants with surface gravities down to log $g$=2.10 (\citet{2006A&A...456.1109M}).  The single 
star which has greatly enhanced [Na/Fe] and [Al/Fe], HIP 114155, is an evolved, metal poor red 
giant with enrichments of 0.53 dex and 0.51 dex, comparable to those found by \citet{2009ApJ...701..837S}.
According to the NLTE correction table of \citet{2003ChJAA...3..316T}, the recommended NLTE correction is
at most -0.10, although the calculations performed do not extend below a temperature of
4500 K.  \citet{1999A&A...350..955G} performed an extensive set of NLTE corrections for Na, and based on their
results, there is a recommended NLTE correction of $\sim$ 0.20 dex.  Even considering these corrections,
the Na abundance remains enhanced.  Although there are few NLTE corrections for Al in the literature, 
\citet{2008A&A...481..481A} suggest NLTE corrections of roughly 0.60 dex upward.  This is opposite to the 
necessary correction to remove the enhancement, however the corrections are for low-metallicities 
([Fe/H]$\approx$-2.00).  Further NLTE calculations for cool, moderately low metallicity giant like
HIP114155 are needed to determine whether the enhanced abundances in this star are a result of NLTE effects.

The other points of interest in Figure \ref{na_al} are the two dwarfs with the greatest
surface gravities ([Na/Fe]=-0.38 in HIP105341 and [Na/Fe]=-0.33 in HIP5027).  Closer
inspection shows that these are the two coolest dwarfs in the sample, perhaps 
pointing to overexcitation/ionization as a culprit for decreased abundances, similar to 
overexcitation/ionization effects observed in cool open cluster dwarfs 
(\citet{2003AJ....125.2085S}, \citet{2004ApJ...603..697Y}, \citet{2005PASP..117..911K}
and \citet{2006AJ....131.1057S}).  

Similar effects are not apparent for [Al/Fe].  A single Na line was measurable with
a relatively low excitation potential of 2.10 eV, while two Al lines of 3.14 eV and 4.02 eV were used.
Additionally, the ionization potential of Al is $\sim$ 0.9 eV higher than for Na.  These differences
are qualitatively consistent with those needed for overexcitation/overionization to be manifest.
This can be further explored by comparing abundances from \ion{Fe}{1} and \ion{Fe}{1}I.   

\subsubsection{Overexcitation and Overionization in Cool Dwarfs: \ion{Fe}{1} and \ion{Fe}{1}I Abundances}

In order to more closely examine the possible effects of overexcitation and overionization for the sample,
abundances have been derived from \ion{Fe}{1} and \ion{Fe}{2} lines using physical surface gravities 
(spectroscopic gravities are unsuitable for this purpose since ionization balance forces agreement between
abundances of \ion{Fe}{1} and \ion{Fe}{2}).
Refer to Figure \ref{overionization_dwarfs} where the difference in abundances between ionized and neutral
Fe are plotted versus temperature.  For stars 
warmer than 4500 K the general trend reveals no overionization within the uncertainties.  The 
same two coolest dwarfs which evince unusually low [Na/Fe], show large degrees of Fe overionization, 

Source of overionization in cool dwarfs are not well-understood, however, one possible explanation is that the stars
are active young dwarfs and, thus, heavily spotted.  Recent work suggests that heavily spotted stars have radii which
are ``puffed'' compared to standard stellar models (\citet{2002ApJ...567.1140T}, \citet{2008A&A...478..507M}.  
An increased radius would decrease the surface gravity of the star compared to unspotted analogs, 
which would result in increased \ion{Fe}{2} line strengths
via overionization.  In order to explore the viability of this explanation,
the radius that corresponds to the surface gravity needed to eliminate the abundance difference between
[\ion{Fe}{2}/H] and [\ion{Fe}{1}/H] was determined for HIP5027.  A 
surface gravity of 3.57 was found to produce agreement between abundances from \ion{Fe}{1} and \ion{Fe}{2}, holding 
temperature and microturbulence constant.  From Yale-Yonsei
isochrones, a mass of 0.66 M$_{\odot}$ is assumed.  The radius for this gravity is R=2.19 R$_{\odot}$.  
The radius corresponding to this mass and the physical surface
gravity of log $g$=4.70 is R=0.60 R$_{\odot}$.  From \citet{2008A&A...478..507M} an upper limit that
can be expected for radius changes in this ``spotted'' regime is $\sim$10\%, well beneath the radius
change implied by the necessary surface gravity change to eliminate overionization and well
outside of the uncertainty in the physical surface gravity.  This points
to a more likely scenario of significant NLTE effects yielding increased overionization as a function
of decreasing temperature as observed in many cool open cluster dwarfs (\citet{2004ApJ...603..697Y},
\citet{2006AJ....131.1057S}).

\subsubsection{Oxygen Abundances: Moving Groups Versus Open Clusters}

Abundances for the $\lambda$7771, $\lambda$7774, 
$\lambda$7775 high excitation potential oxygen triplet have been derived from equivalent widths.  
Since abundances derived from the triplet are believed to be enhanced by NLTE effects,
corrections from the work of \citet{2003A&A...402..343T} have been applied
to derive NLTE corrected abundances from the triplet lines.
The equivalent widths
for the triplet, the LTE oxygen abundances, and the final NLTE oxygen abundances are shown
in Table \ref{oxygen}.

The $\lambda$7774 {\AA} and $\lambda$7775 {\AA} lines appear 
to be enhanced as a general function of decreasing temperature in both dwarfs 
(Figure \ref{oxygen_dwarf_diff}) and giants.  A similar 
enhancement of the central line (7774.1 {\AA}) in Hyades giants was noted by 
\citet{2006AJ....131.1057S}.  They believed this 
enhancement to be due to a possible blend with an \ion{Fe}{1} feature at 7774.00 {\AA}.  
While the nature of any blending for the reddest feature (7775 \AA) is unclear, 
visual inspection of the spectral line reveals a slight asymmetry, possibly indicating a blend.
The distinct increase in [O/H] abundances derived from the red features of the triplet as
a function of decreasing temperature suggest that only the blue line (7771.1 {\AA})
of the triplet should be used for oxygen abundance determinations in cooler stars. 

In order to test the possibility of an Fe blend as discussed
above, two cool stars of the sample with no measurable oxygen abundances (HIP5027 and HIP105341)
were examined to see if they showed any indications of an Fe blending feature near 7774 {\AA}.  
In HIP5027 a possible detection of a feature at 7774 {\AA} was found to have a measured 
equivalent width of roughly 6.0 m{\AA}.  This strength is not inconsistent with the expected 
contribution required from two nearby \ion{Fe}{1} features at 7773.979 {\AA} and 
7774.06 {\AA} for the derived Fe abundance.

Neglecting the two red triplet lines in the cool dwarfs, the [O/H] trend of the our dwarf 
sample is plotted along with the Pleiades trend from \citet{2004ApJ...602L.117S} 
(where [Fe/H]=0.00 was assumed to calculate [O/Fe]), and the Hyades trend of 
\citet{2006AJ....131.1057S} (where [Fe/H]=$+$0.13 was assumed to calculate [O/Fe])
in Figure \ref{oxygen_dwarf_simon}.  
Using $\lambda$7772 triplet-based [O/H] abundances in 45 Hyades dwarfs, they found 
a remarkable increase in [O/H] as a function of decreasing temperature for stars with 
T$_{eff}$$\le$5400 K.  The increase of [O/H] in the $\sim$ 120 Myr old Pleiades appeared
to be steeper than that in the $\sim$ 625 Myr old Hyades, 
perhaps pointing to an age-related effect whereby [O/H] enhancements in cooler
stars decrease as a function of increasing age.  

Our field dwarfs do not show a drastic increase in abundance as a function of decreasing temperature.
The single star
that appears to reside within the increasing Hyades trend at cooler temperatures is metal weak
(HIP 42499, [Fe/H]=-0.56), resulting in [O/Fe]=+0.47.  The enhanced [O/Fe] ratio at this low metallicity 
is unsurprising and coincides with the characteristic field dwarf enhancements observed as a function 
of decreasing temperature for oxygen in other metal poor field stars \citep{1989ApJ...347..186A}.
If the abundance trend observed by \citet{2004ApJ...602L.117S} and \citet{2006AJ....131.1057S} 
is age dependent, the lack of a distinct trend of increasing [O/Fe] with decreasing abundance
may point to the stars in the sample being older than the Hyades,
not inconsistent with the 2.7 Gyr age of the dominant subsample identified above.  
If not an age-related effect, then an as yet unknown dichotomy between oxygen 
abundances in field stars and cluster stars would have to be explored with 
abundances of field stars of quantifiable age.  

For the giant stars in the sample, oxygen abundances have been derived from the infrared triplet
and from the forbidden line at $\lambda$6300.301 {\AA}, through use of the \emph{blends} driver
of MOOG, following the approach of \citet{2006AJ....131.1057S}.  

In examining the giant triplet abundances, a similar effect as in the dwarfs is observed as temperatures
decrease with enhancements in oxygen abundances derived from both the 7774 {\AA} and
7775 {\AA} lines.  
NLTE corrections were applied to the $\lambda$7771 triplet abundances by interpolating within the grids
of \citet{2003A&A...402..343T}.  The results of these corrections are presented in  
Figure \ref{giant_oxygen_forbidden}
where forbidden minus permitted [O/H] differences versus temperature are plotted.  Notice that as the 
temperature decreases, the
abundance from the redder lines of the triplet appear to be enhanced relative to the forbidden
line.  While the NLTE corrections decreased the abundance enhancements in the cooler stars of the sample, 
they did not eliminate them.  This yields further evidence of blending
effects in the reddest lines of the triplet as a function of coolier temperature.  
For the purposes of this paper, the oxygen abundances derived from the red 
features of the triplet will not be used.

In Figure \ref{giant_forbid_bluetriple} the difference in abundance from the forbidden
oxygen line (6300.34 {\AA}) and the NLTE corrected blue triplet line (7771 {\AA}) is plotted versus
temperature (top plot) and surface gravity (bottom plot).  The dotted line shows a zero difference
between the two abundances.  The NLTE-corrected permitted oxygen abundances (7771 {\AA})
appear to agree well with the forbidden oxygen abundance (6300 {\AA}) indicating that the blue line
of the triplet can provide a reliable oxygen abundance when proper care is taken to make the
necessary NLTE corrections.  

The single outlier is the highly evolved giant HIP 114155.  The larger abundance 
from the blue triplet feature
in this star is believed to be from NLTE effects that are not removed using the
corrections of \citet{2003A&A...402..343T} as the grid for the corrections does not
extend below 4500 K.  While the temperature extrapolation is sufficient for less evolved stars
(i.e. NLTE triplet abundances in stars with surface gravities above 2.0 all agree with the 
forbidden abundance, even at temperatures below 4500 K), the corrections for more
evolved stars, with surface gravities $\sim$1.00, are significantly larger. 
The good agreement between all other forbidden and blue triplet oxygen abundances indicates 
the inadequacy extrapolating the NLTE corrections in cool, evolved stars.  

The final salient point to make regarding the oxygen abundances is to address the alleged
CN blending feature previously discussed.  As mentioned, \citet{1963rspx.book.....D}
list CN features at 6300.265 {\AA} and two features at 6300.482 {\AA} with gf
values of 5.78E-3, 6.82E-3 and 2.01E-3.  Recall that the inability to adequately calibrate a linelist
including these features with a high resolution atlas of Arcturus led to the
features not being utilized in the derivation of forbidden line oxygen abundances.  With the good
agreement between forbidden oxygen neglecting the CN features and the NLTE corrected
blue line of the triplet in Figure \ref{giant_forbid_bluetriple}, it is suggested that 
the CN blending features may not be important.

\section{SUMMARY}

The existence of spatially unassociated groups of stars moving through the solar neighborhood
with common U and V kinematics has been explored for over half a century 
\citep{1958MNRAS.118...65E}.  Despite this long history, the exact origins of these
so called moving groups is still a matter of some debate.  The classical view contends
that they are dissolved open clusters which have retained common kinematics and 
drifted into spatially elongated stellar streams.  If this is indeed true, moving group
members should possess similar characteristics to those of open cluster stars:
particularly, common chemical abundances and residence along a distinct evolutionary sequence in an
HR diagram. 

In order to address the viability of moving groups being dissolved open clusters, 
we have performed a high resolution spectroscopic abundance analysis
of a 34 star sample of the kinematically distinct Wolf 630 moving group, selected for
its residence in a sparsely populated region of the \emph{UV} plane in the solar
neighborhood.  Our abundance measurements reveal that the sample can not be characterized 
by a uniform abundance pattern.  The individual
stars have been closely scrutinized, making use of abundances, evolutionary 
state and qualitative age information to constrain membership as an unlikely,
possible or likely member of a subsample with a dominant abundance trend
and consistent age.  There
appears to be a group with a weighted mean of [Fe/H]=-0.01 $\pm$ 0.02 
(uncertainty in the weighted mean) that is composed of 19 stars.  
These final members are well traced by an evolutionary sequence 
of 2.7 $\pm$ 0.5 Gyr as determined from Yale-Yonsei isochrones 
\citep{2004ApJS..155..667D}.  Thus, the existence of moving groups
as relic structures of dissolved clusters remains plausible based on the homogeneity 
of the subgroup identified above.  

We have also explored some of the additional uses for abundances in moving groups in 
chemically tagging the galactic disk.  We found evidence for overexcitation/overionization
effects from both Na and from \ion{Fe}{1} versus \ion{Fe}{2} abundances in the coolest dwarfs of the sample,
likely attributable to increasing NLTE effects as a function of cooling temperature.  We find
the necessity to apply NLTE corrections of 0.10-0.20 dex to Na abundances in giant stars.
Finally, we derived oxygen abundances for the stars in the sample from both the
forbidden line at 6300 {\AA} and the near-IR triplet.  First, we find evidence for blending in the IR
triplet in both dwarfs and giant stars, possibly by \ion{Fe}{1} features near the $\lambda$7774 line.  
Second, we find that NLTE effects on \ion{O}{1} in low log $g$ cool giants are important and cannot be accounted
for by extrapolating current NLTE calculations.  Finally, we find reliable oxygen abundances
from the forbidden line in giant stars and again find evidence of increased NLTE effects
as a function of cooling temperature manifested in increased triplet derived abundances.

\begin{acknowledgements}
The authors would like to gratefully acknowledgement support for this work provided by 
NSF grants AST-0908342 and AST-0239518.  Furthermore,
we would like to thank the referee for many useful comments which place the work into a broader context.
\end{acknowledgements}

\clearpage

\begin{figure}
\plotone{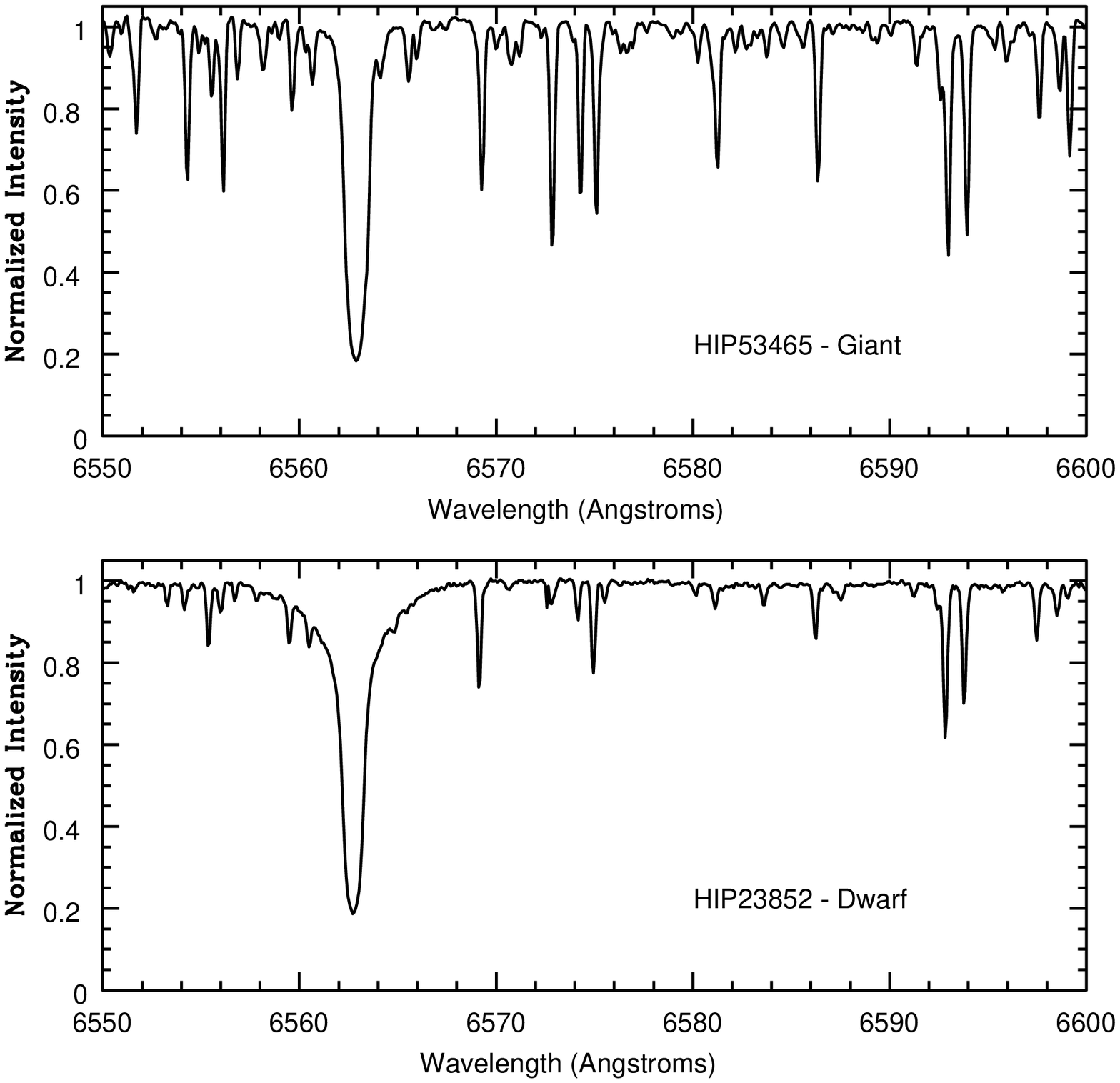}
\caption[Sample Spectra]{Sample normalized spectra of our 34 star sample.  The top panel 
shows a giant star and the bottom
displays a dwarf.  The typical continuum level S/N in these spectra are $\sim$200.}\label{spectra}
\end{figure}

\begin{figure}
\plotone{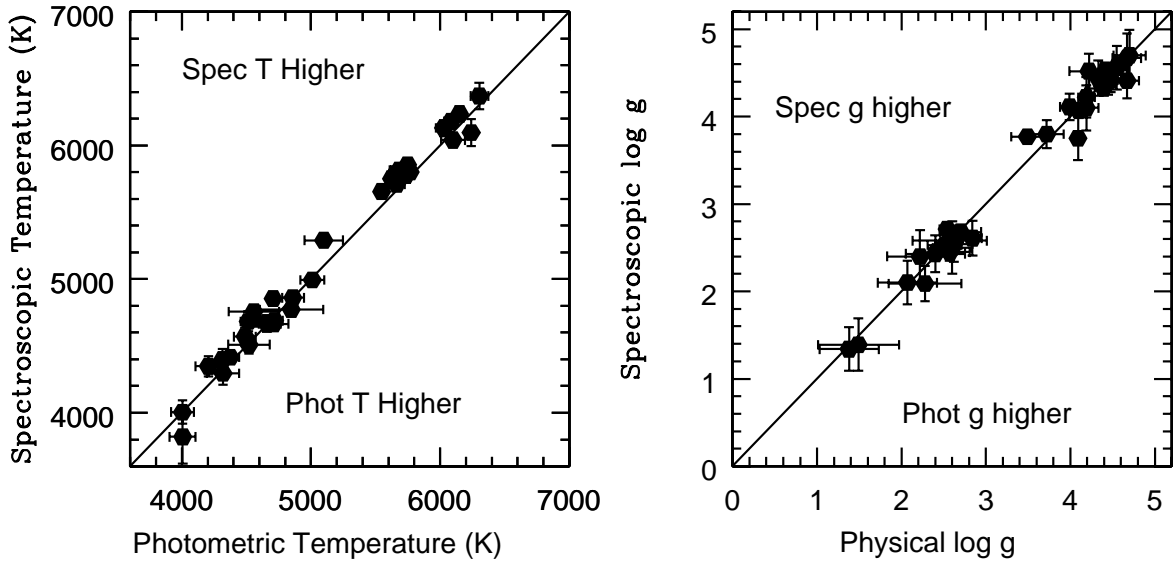}
\caption[Spectroscopic Versus Photometric Parameter Differences]{The
spectroscopic temperatures are plotted versus photometric temperatures in the 
left plot and the spectroscopic gravities versus physical gravities are plotted
in the right plot.
The line is plotted to show perfect agreement between the two values.
The differences between the spectroscopic and photometric parameters agree within the 
uncertainties in the respective mean differences.}\label{param_diff}
\end{figure}		

\begin{figure}
\plotone{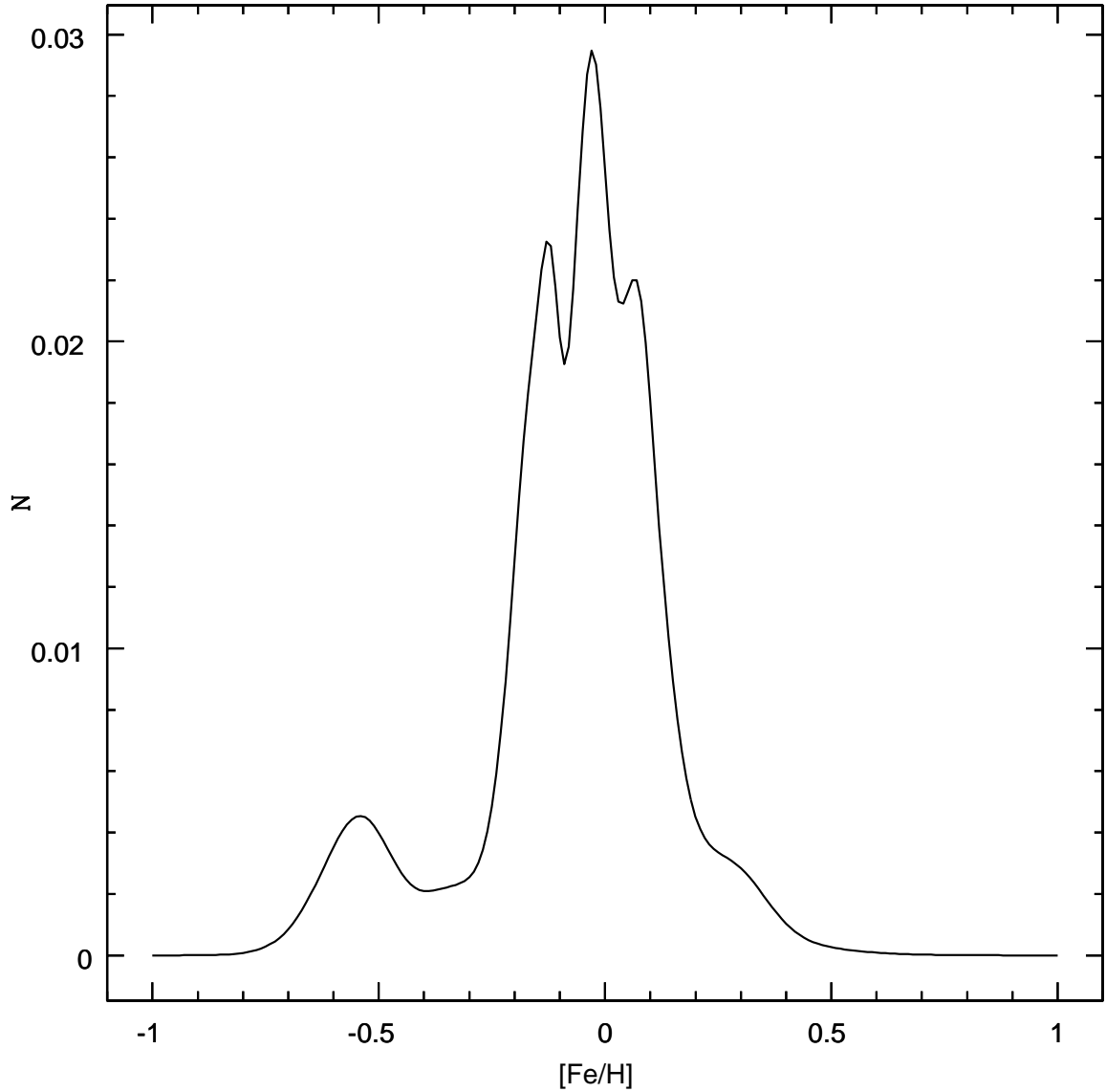}
\caption[Smoothed histogram of Fe/H]{The metallicities of our 34 stars are plotted as guassians with
central peaks at a given star's metallicity and $\sigma$ equal to the uncertainty in the [Fe/H].
The guassians are normalized to unit area and summed to yield the smoothed abundance histogram.
The peak at [Fe/H]$\sim$-0.50 is from 3 low metallicity stars and the bump at [Fe/H]$\sim$0.30 is
from 2 high metallicity stars.}\label{fe_smooth}
\end{figure}

\begin{figure}
\plotone{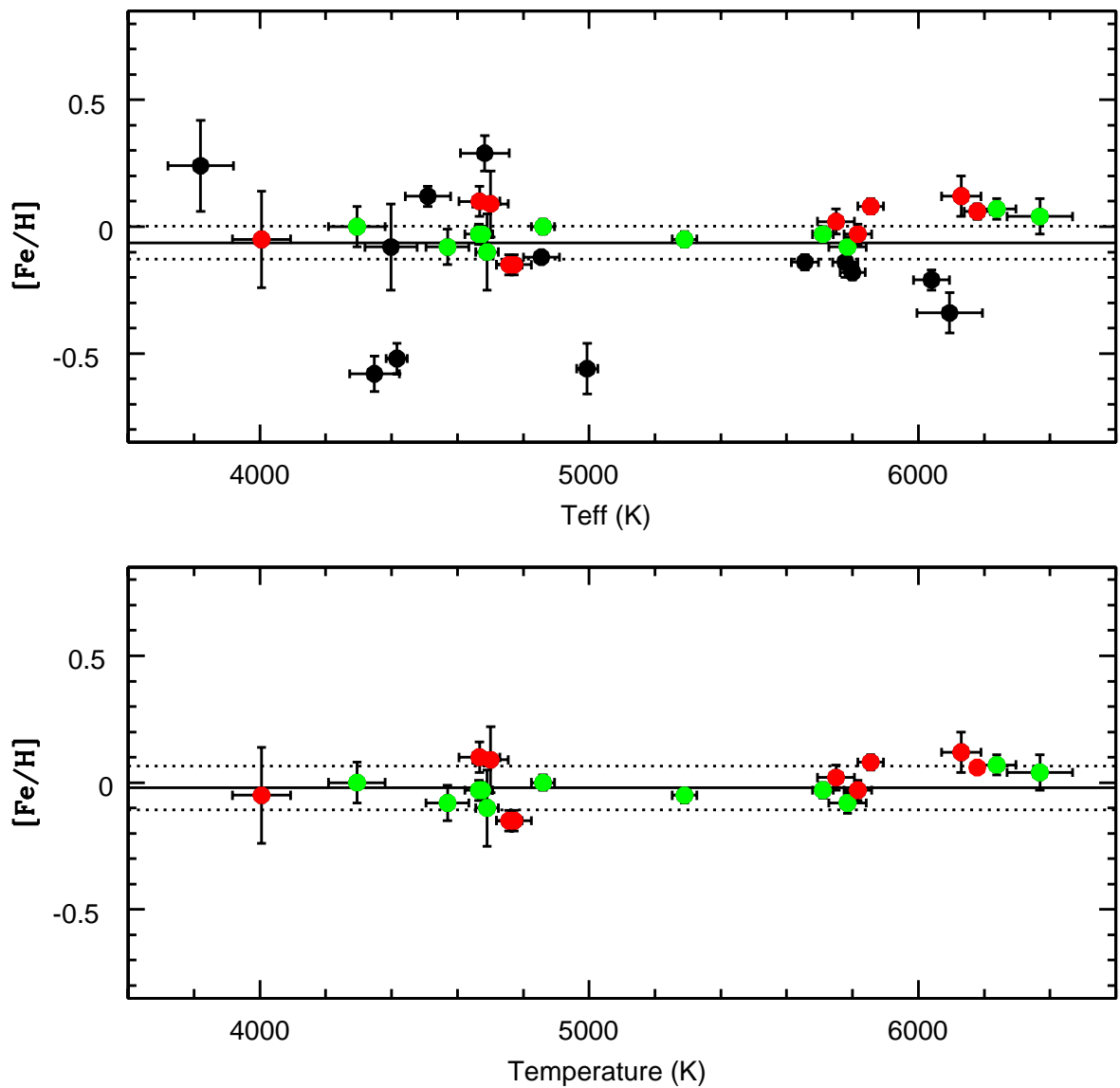}
\caption[Metallicity Bands for Fe/H]{The [Fe/H] is plotted versus temperature for the full sample
of stars (top) and the possible (red) and likely (green) homogeneous members (bottom).  The solid line gives the weighted mean of the sample
while the dotted lines are 3-$\sigma$ deviations from this mean.  If a star rests
within the dotted lines (i.e. the abundance band) within its respective uncertainty,
then it is considered homogeneous with the dominant sample.  Those stars which
rest far outside the abundance band in the full sample plot are iteratively removed
as unlikely members until convergence to a dominant abundance is achieved, as seen in the bottom plot.}\label{fe_abtemp}
\end{figure}

\begin{figure}
\plotone{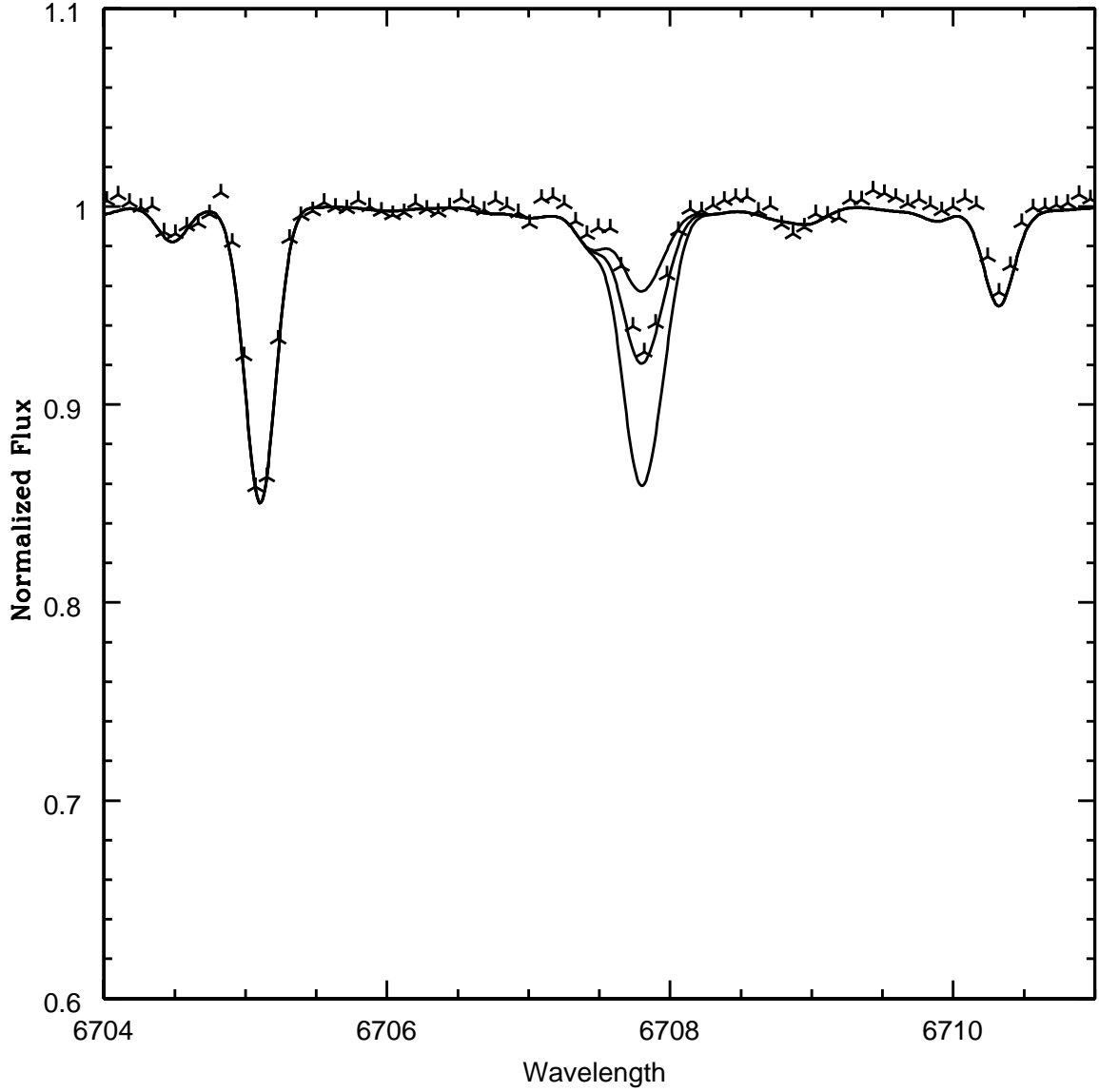}
\caption[Lithium Synthesis]{Sample lithium synthesis for HIP 23852.  The crosses are the
observed spectrum while the lines are lithium abundances of logN(Li)=2.30, 2.00 (best fit)
and 1.97.}\label{lithium_synthesis}
\end{figure}

\begin{figure}
\plotone{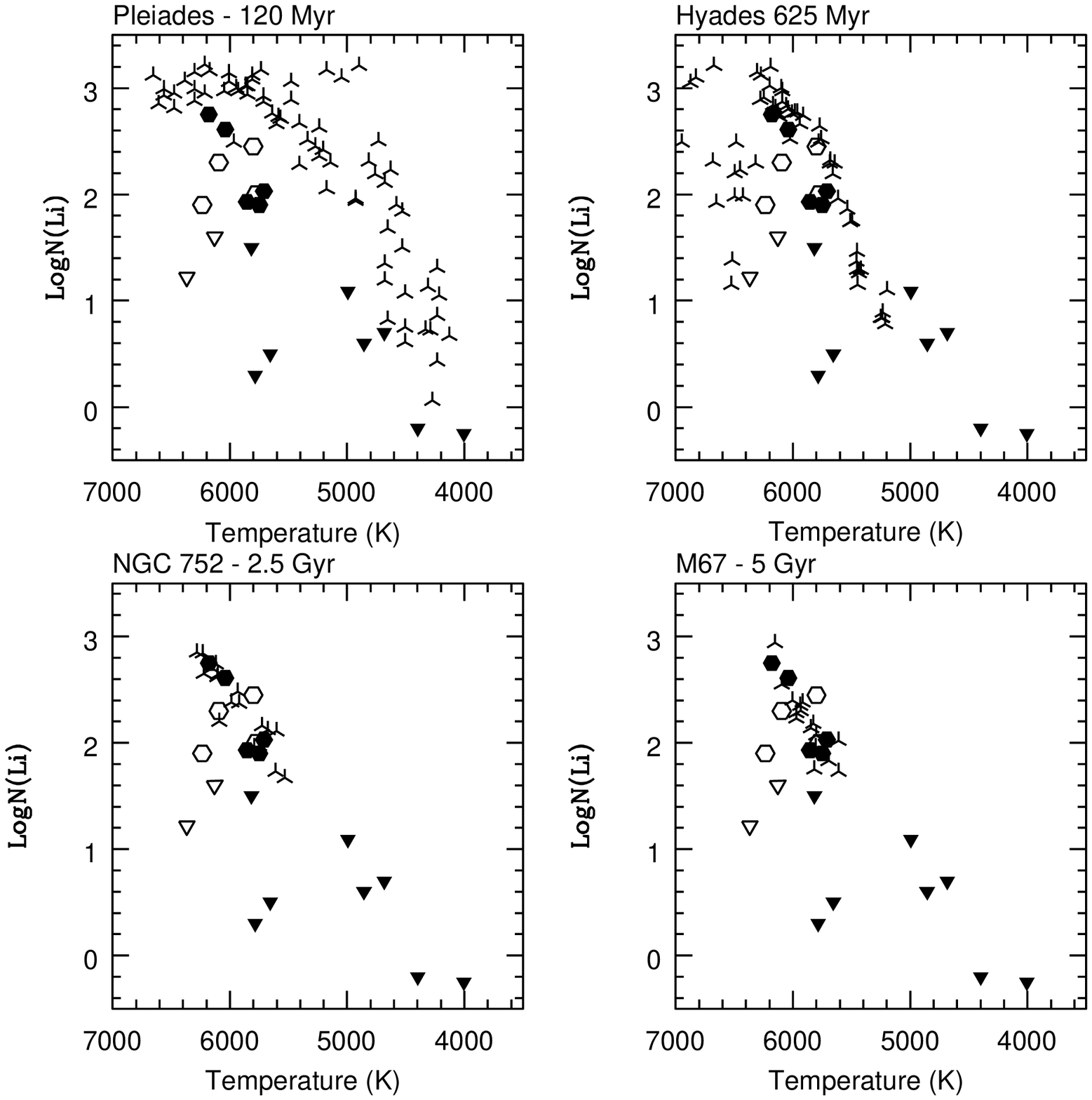}
\caption[Lithium Abundances]{Lithium abundances for the Pleiades (top left-\citet{2000AJ....119..859K}),
the Hyades (top right-\citet{1995ApJ...446..203B}, NGC752 (bottom left-\citet{2004A&A...426..809S}) and
M67 (bottom right-\citet{1999AJ....117..330J} (plotted as crosses) and our Wolf 630 candidates.  Filled
hexagons are for dwarfs, filled triangles are upper limits for dwarfs, open hexagons are for subgiants open triangles 
are upper limits for subgiants.  Specific abundances for individual stars are discussed in more 
detail in the text.}\label{lithium}
\end{figure}

\begin{figure}
\plotone{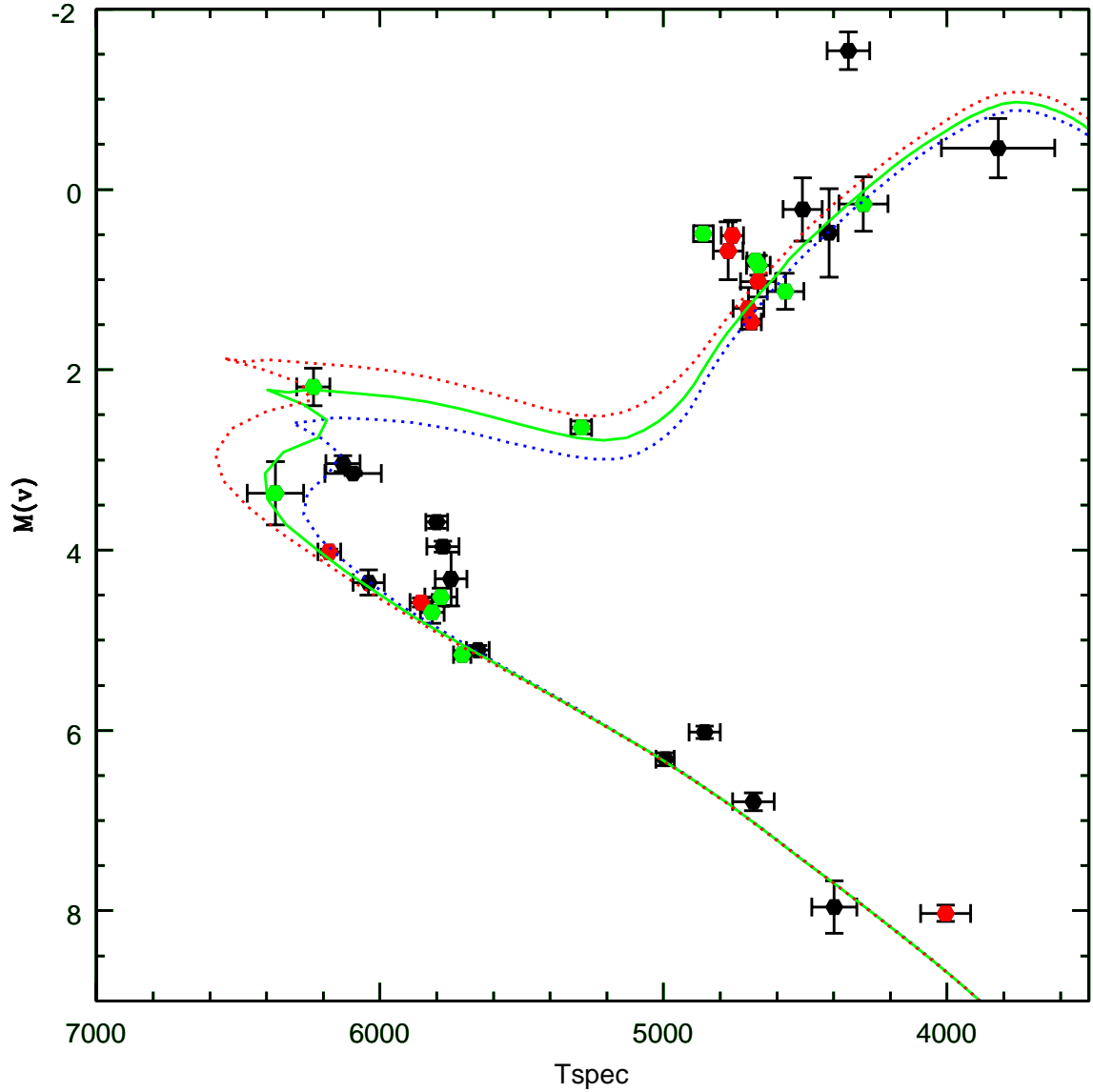}
\caption[Final Spectroscopic HR Diagram: A Kinematic-Chemical Sample]{The
HR diagram of the final candidate members of a common chemical group
with the distinct UV kinematics of the classical Wolf 630 group.  Green
points are likely members while red points are possible members.  Unlikely
members are plotted as black points.  Yale-Yonsei isochrones of 2.2, 2.7 and 3.2 Gyr
iare shown.}
\label{HR_spec_final}
\end{figure}

\begin{figure}
\plotone{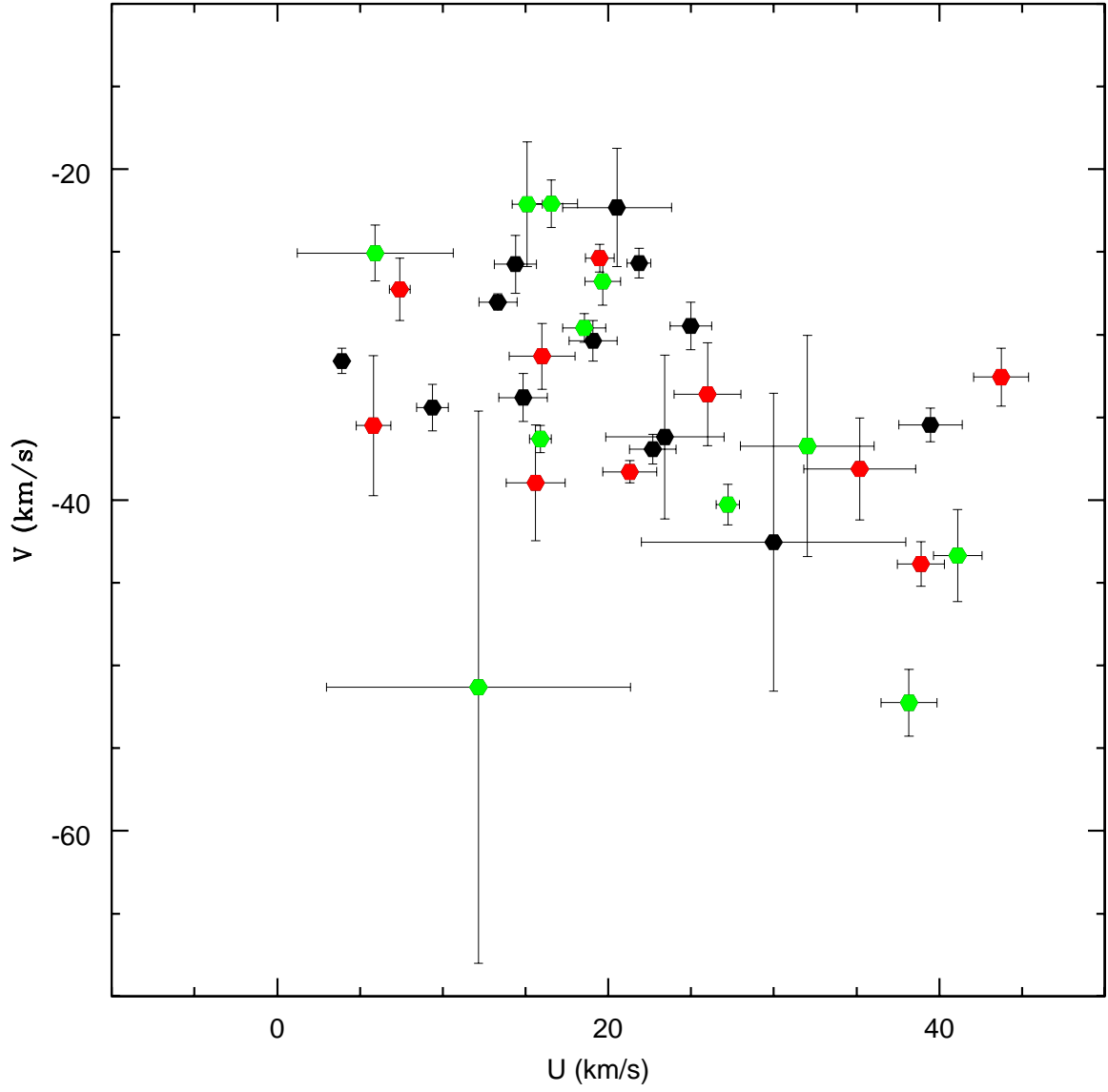}
\caption[Final UV Kinematics: Kine-Chemical Sample]{Plot 
of the U and V kinematics for the sample with likely members
plotted in red and possible members in green.  Black points are non-members}\label{uvw_final}
\end{figure}

\begin{figure}
\plotone{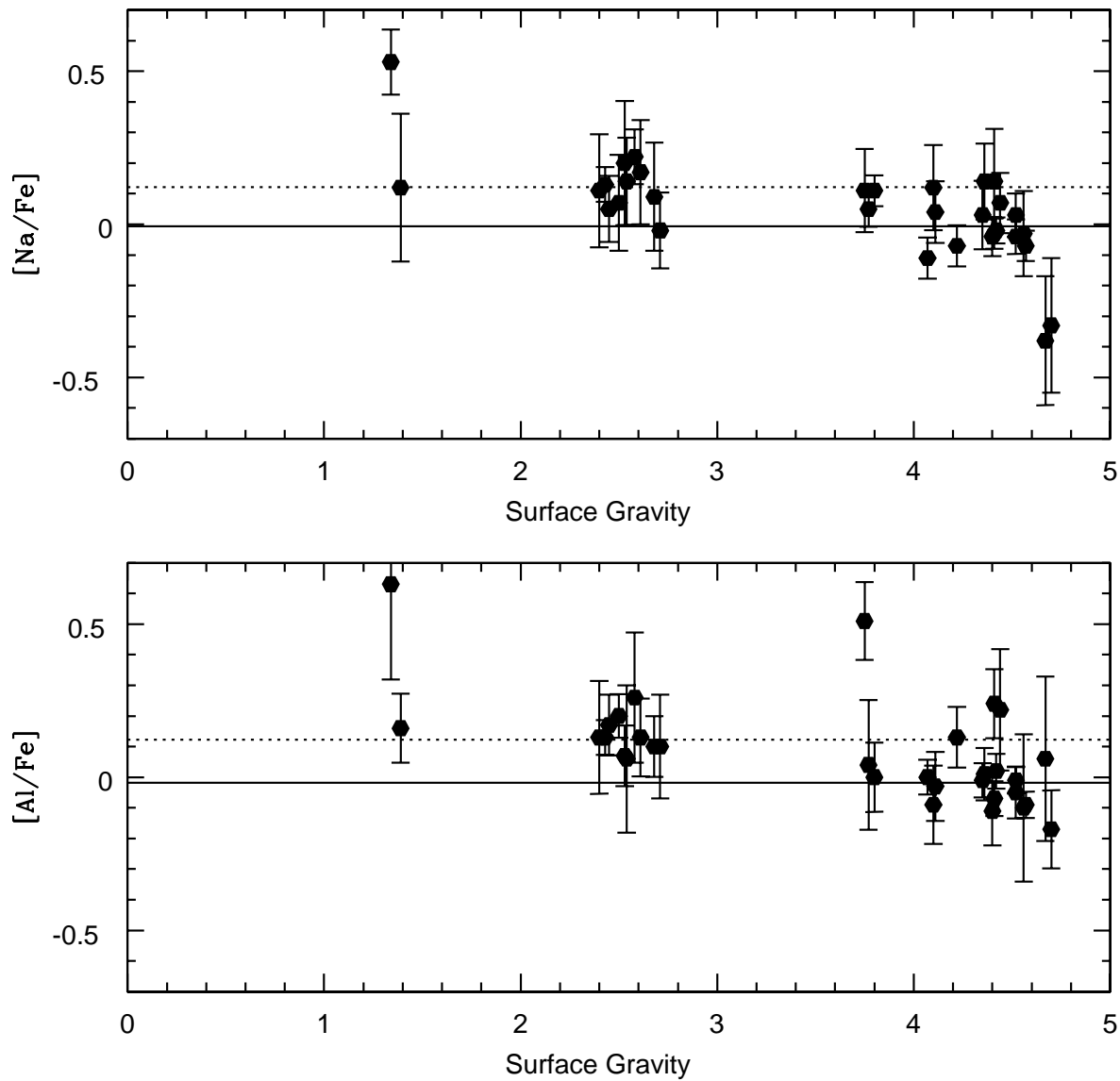}
\caption[Na/Fe and Al/Fe Abundances Versus Surface Gravity]{The abundances
[Na/Fe] (top) and [Al/Fe] (bottom) for all stars with measurable lines of Na and/or Al are plotted 
versus surface gravity.  The solid line
gives the weighted mean [X/Fe] for the dwarfs, neglecting the two with unusually
low [Na/Fe].  The dotted line gives the weighted mean [X/Fe] for the
subgiants and giants, neglecting the giant with unusally high [Na/Fe]
and [Al/Fe].  Subgiant and giant abundances are $\sim$ 0.10 dex higher
than dwarfs.}\label{na_al}
\end{figure}

\begin{figure}
\plotone{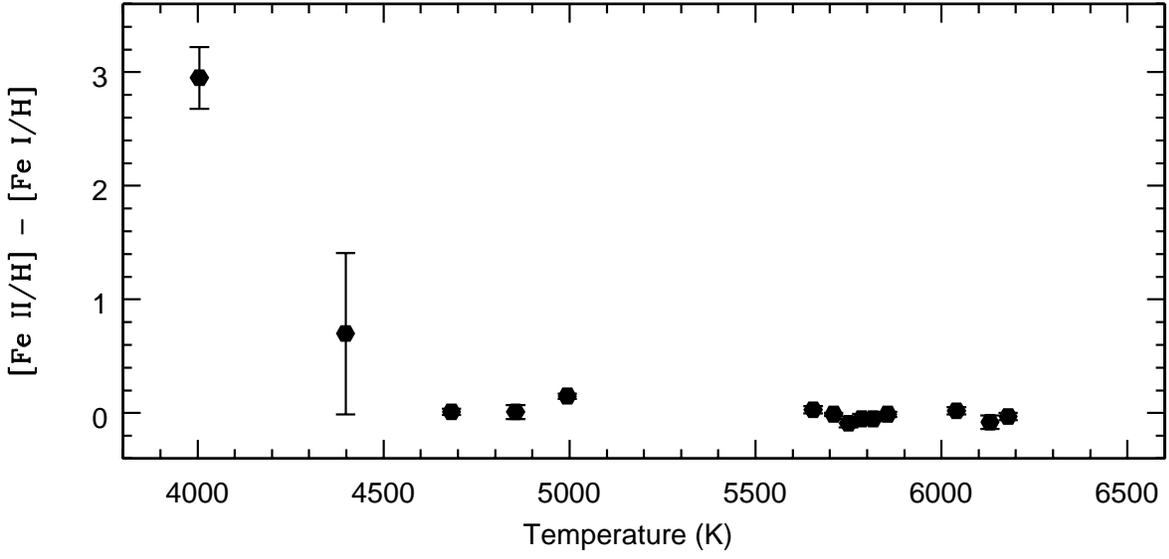}
\caption[Overionization in Cool Dwarfs]{The difference [\ion{Fe}{1}I/H]-[\ion{Fe}{1}/H] is plotted
versus temperature.  Notice the clear overionization
in the two coolest dwarfs of the sample.}\label{overionization_dwarfs}
\end{figure}

\begin{figure}
\plotone{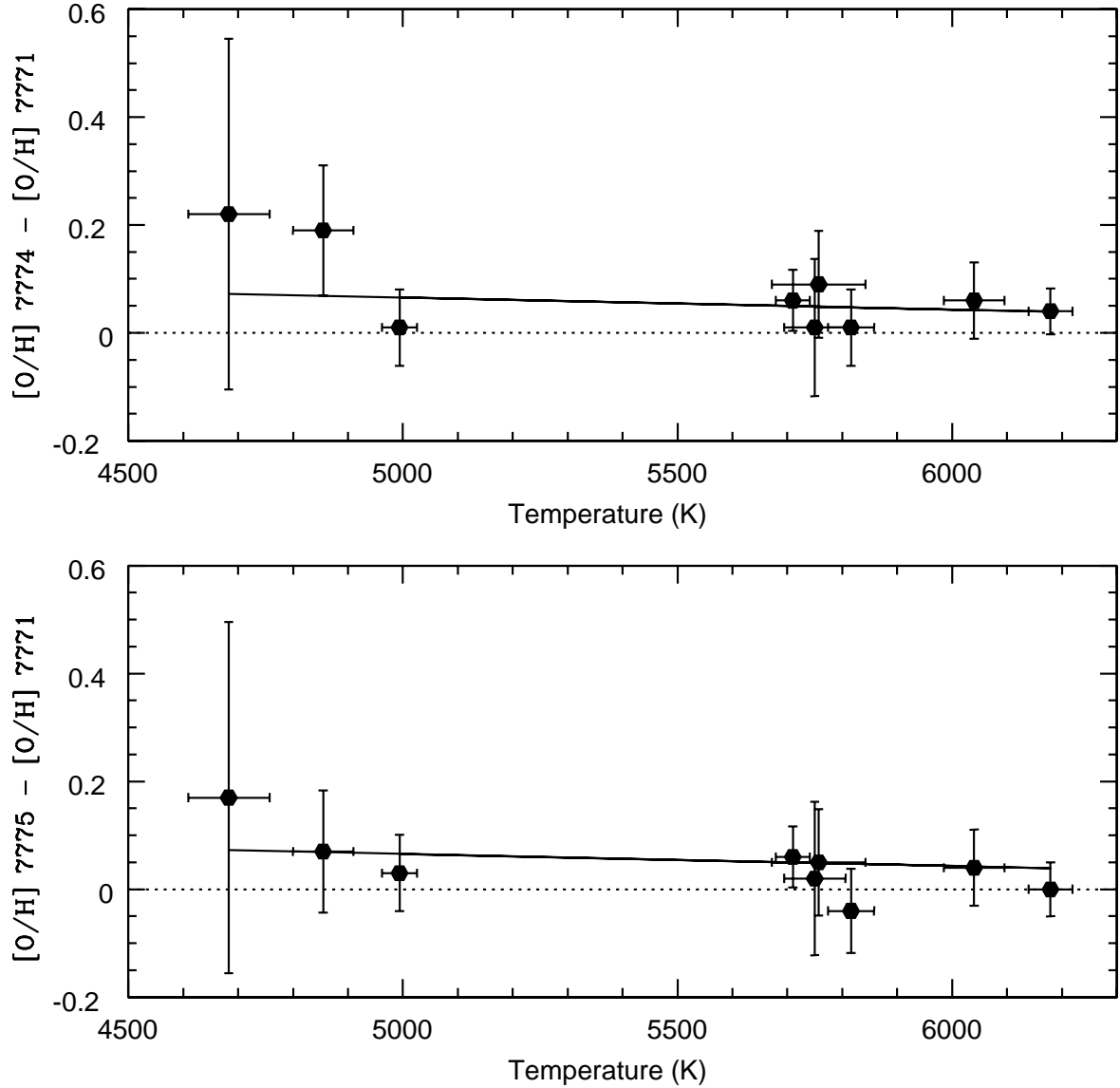}
\caption[Differences in Triplet Oxygen Abundances: Dwarfs]{Differences in 
oxygen abundances
for dwarf stars derived from the infrared triplet.  The top plot shows the 
difference in the abundance from the 7774 line and the 
7771 line.  The difference in abundance between 
these two lines for the coolest two stars in the sample is of order 0.20 dex.  
The difference between the 7775 line and the 7771 line is slightly more modest,
but the general trend is for the cooler stars to yield slight abundance enhancements.}\label{oxygen_dwarf_diff}
\end{figure}

\begin{figure}
\plotone{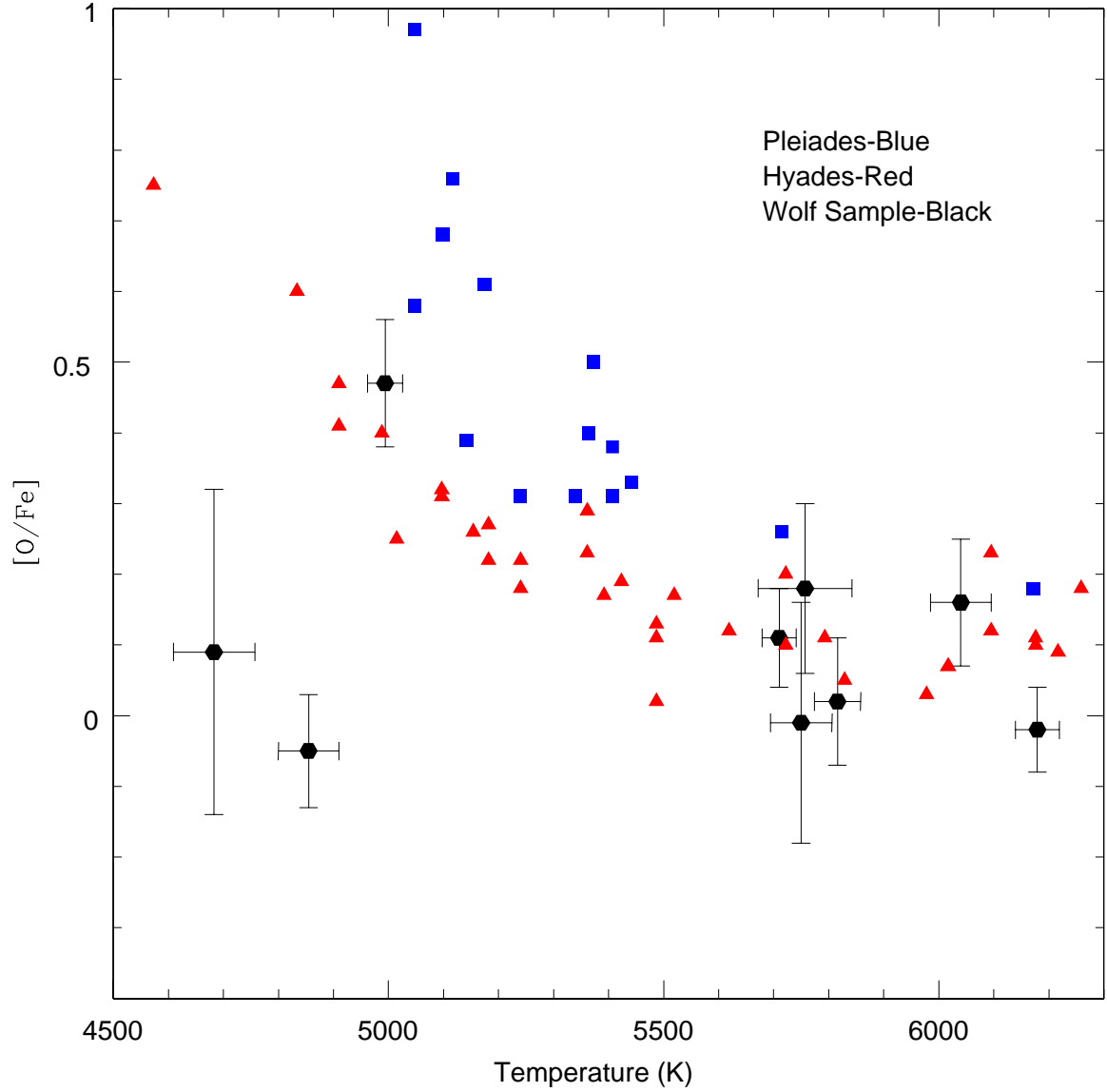}
\caption[O/H Versus Temperature: Dwarfs]{Oxygen abundances [O/Fe] versus temperature for 
the Wolf 630 sample that were determined to be chemically homogeneous (black), the Pleiades (blue) and the Hyades (red).}
\label{oxygen_dwarf_simon}
\end{figure}

\begin{figure}
\plotone{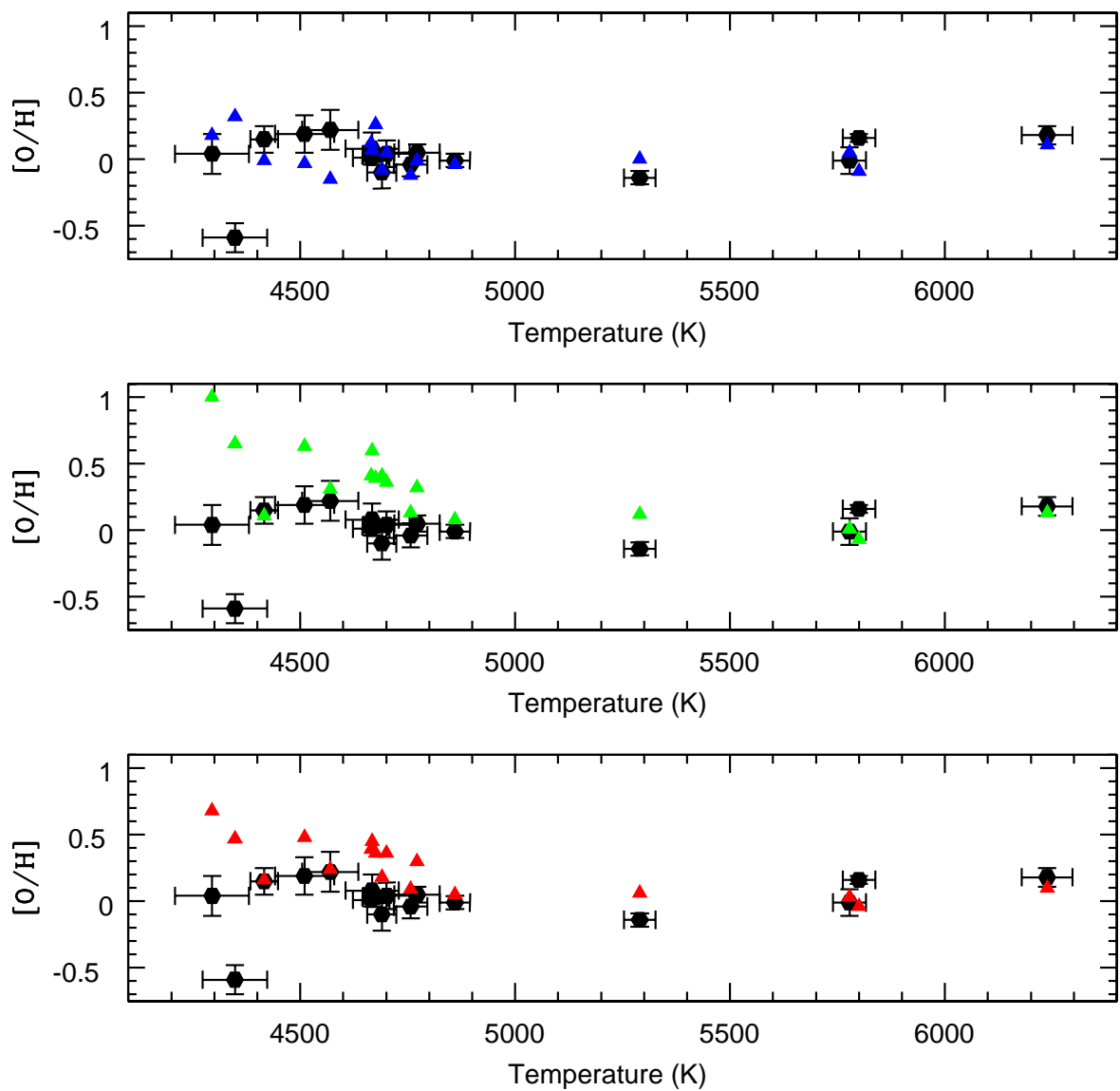}
\caption[Giant Star O/H from Forbidden Line vs. Triplet Abundances]{The
[O/H] abundance from the forbidden line (black hexagons) for subgiant and giant 
stars is plotted versus				 
temperature in all windows.  The top plot shows the NLTE [O/H]			
abundances from the 7771 line of the triplet (blue triangles),  				
the middle plot shows the NLTE [O/H] from the 7774 line					
of the triplet (green triangles) and the bottom plot gives NLTE 			
[O/H] from the 7775 line (red triangles).  The abundances derived					
from the 7771 line agree well, after NLTE corrections, with abundances from 			
the forbidden line, but [O/H] abundances from			
the redder lines of the triplet increase as a function of decreasing temperature.}\label{giant_oxygen_forbidden}
\end{figure}	

\begin{figure}
\plotone{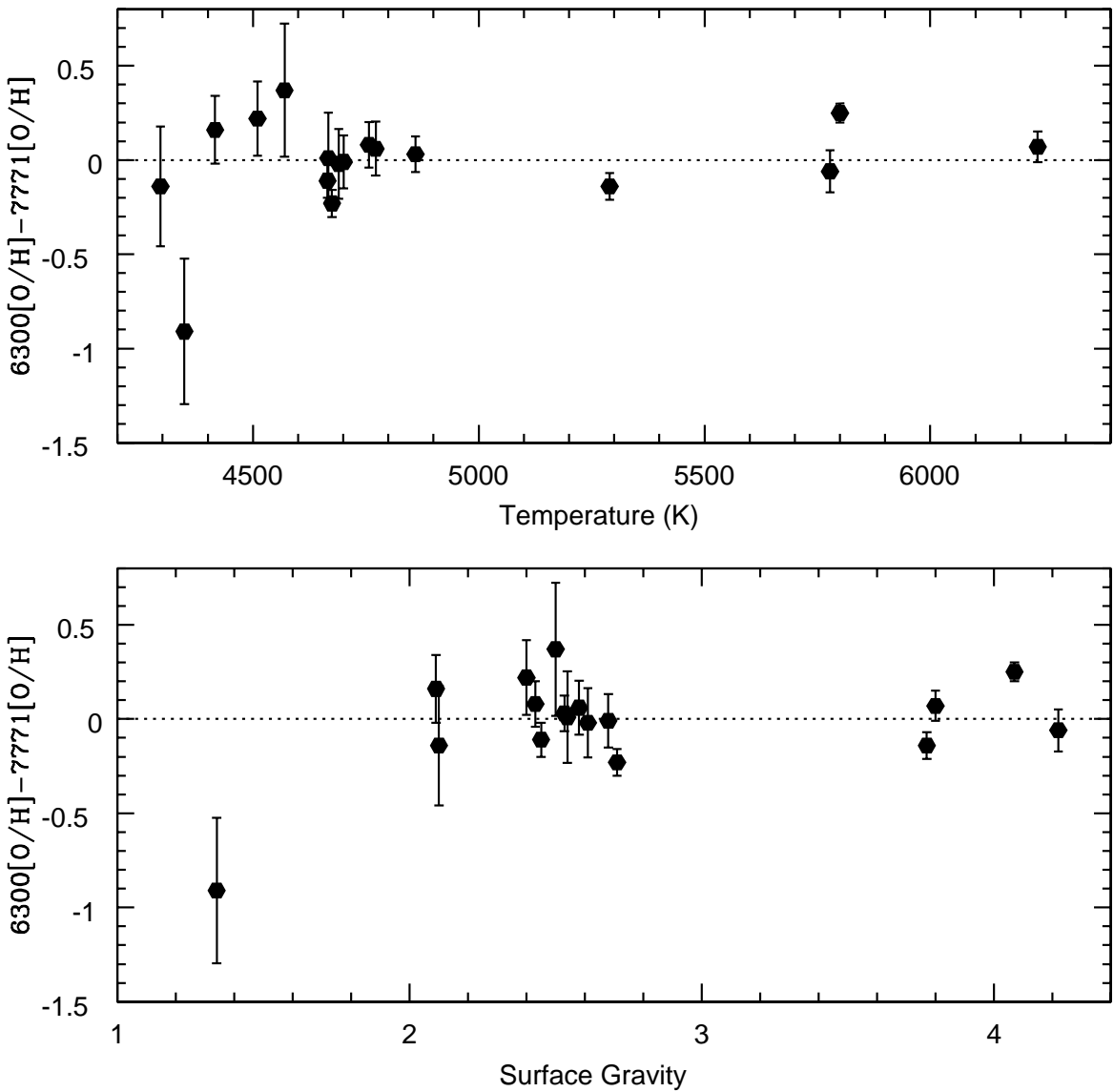}
\caption[Giant O/H From Forbidden vs. 7771 Line]{Differences in 
oxygen abundance between the forbidden line at 6300 {\AA} and 7771 {\AA} for the giant
and subgiant stars.  The NLTE
corrected abundances from the triplet line generally agree with 
the 6300 forbidden line, with the exception of the metal weak cool star, HIP 114155.
The clear agreement between the forbidden and the NLTE triplet abundance until reaching
a low surface gravity possibly indicates that greater than expected NLTE effects
impact the triplet abundances in more evolved stars.}
\label{giant_forbid_bluetriple}
\end{figure}

\begin{deluxetable}{l c c c c c c c}
\rotate
\tablewidth{0 pt}
\tablecaption{Kinematic Information}
\tablehead{
\multicolumn{1}{l}{HIP}   &
\multicolumn{1}{c}{$\pi$}       &    
\multicolumn{1}{c}{PM RA}               &
\multicolumn{1}{c}{ PM DEC }           &
\multicolumn{1}{c}{Radial Velocity } &
\multicolumn{1}{c}{ U }              &
\multicolumn{1}{c}{ V }              &
\multicolumn{1}{c}{ W }            \\
\multicolumn{1}{l}{}&
\multicolumn{1}{c}{(mas)}&            
\multicolumn{1}{c}{(mas/yr)}    &          
\multicolumn{1}{c}{(mas/yr)}     &     
\multicolumn{1}{c}{(kms$^{-1}$)}  &     
\multicolumn{1}{c}{(kms$^{-1}$) }  &      
\multicolumn{1}{c}{(kms$^{-1}$) }   &     
\multicolumn{1}{c}{(kms$^{-1}$) }    
}
\startdata
102531 & 31.69 $\pm$ 2.37 &   -6.10  $\pm$  2.88 & -201.74 $\pm$ 1.84 &  -7.0  $\pm$ 0.9 & 16.6  $\pm$ 1.6  & -22.1  $\pm$ 1.4  & -14.1  $\pm$ 1.3 \\
102532 & 32.14 $\pm$ 1.19 &  -25.88  $\pm$  1.33 & -196.27 $\pm$ 0.8  &  -6.5  $\pm$ 0.9 & 17.8  $\pm$ 0.9  & -21.2  $\pm$ 1.0  & -11.3  $\pm$ 0.6 \\
103983 & 15.70 $\pm$ 1.29 &  -71.25  $\pm$  1.53 & -155.22 $\pm$ 0.56 &  -6.8  $\pm$ 2.0 & 35.2  $\pm$ 3.4  & -38.1  $\pm$ 3.1  & -3. 8  $\pm$ 1.1 \\
104521 & 28.38 $\pm$ 0.90 &   49.07  $\pm$  0.90 & -151.85 $\pm$ 0.64 &  -17.0 $\pm$ 0.9 &  1.2  $\pm$ 0.5  & -29.0  $\pm$ 0.9  & -12.5  $\pm$ 0.7 \\
105341 & 61.04 $\pm$ 1.31 & -173.35  $\pm$  1.46 & -721.75 $\pm$ 0.72 &  21.1  $\pm$ 2.0 & 38.9  $\pm$ 1.4  & -43.9  $\pm$ 1.3  & -18.2  $\pm$ 1.3 \\
11033  &  6.47 $\pm$ 0.87 &    9.02  $\pm$  0.73 & -58.12  $\pm$ 0.57 & -3.0   $\pm$ 5.0 & 23.4  $\pm$ 3.6  & -36.2  $\pm$ 4.9  & -2.9   $\pm$ 4.7 \\
112222 & 24.86 $\pm$ 0.93 &    9.80  $\pm$ 10.66 & -171.17 $\pm$ 1.25 &  -2.0  $\pm$ 5.0 & 15.1  $\pm$ 0.9  & -22.1  $\pm$ 3.8  & -18.9  $\pm$ 3.4 \\
112447 & 61.54 $\pm$ 0.77 &  233.06  $\pm$  0.80 & -492.04 $\pm$ 0.61 &  -5.3  $\pm$ 0.9 &  3.9  $\pm$ 0.1  & -31.6  $\pm$ 0.8  & -27.8  $\pm$ 0.7 \\
\enddata
\label{uvw}
\end{deluxetable}
\begin{deluxetable}{l c c c c}
\tablewidth{0 pt}
\tablecaption{Solar Equivalent Widths}
\tablehead{
\multicolumn{1}{l}{Wavelength}& 
\multicolumn{1}{c}{Ion}&
\multicolumn{1}{c}{Excitation Potential }&
\multicolumn{1}{c}{log(gf) }& 
\multicolumn{1}{c}{Equivalent Width } \\
\multicolumn{1}{l}{\AA       }       &
\multicolumn{1}{c}{} &
\multicolumn{1}{c}{eV}       &
\multicolumn{1}{c}{  }&
\multicolumn{1}{c}{m\AA}
}
\startdata
  5505.881 & \ion{Fe}{1} &        4.42          & -1.30   &           54.7   \\
  5506.778 & \ion{Fe}{1} &        0.99          & -2.80   &          118.6   \\
  5522.447 & \ion{Fe}{1} &        4.21          & -1.55   &           41.4   \\
  5536.580 & \ion{Fe}{1} &        2.83          & -3.81   &            8.1   \\
  5539.280 & \ion{Fe}{1} &        3.64          & -2.66   &           17.5   \\
  5543.936 & \ion{Fe}{1} &        4.22          & -1.14   &           61.8   \\
  5546.500 & \ion{Fe}{1} &        4.37          & -1.31   &           49.9   \\
  5546.991 & \ion{Fe}{1} &        4.22          & -1.91   &           26.5   \\
\enddata
\label{eqw}
\end{deluxetable}                                            
\begin{deluxetable}{l c c c c}
\tablewidth{0 pt}
\tablecaption{Basic Physical Parameters}
\tablehead{
\multicolumn{1}{l}{HIP} &
\multicolumn{1}{c}{T$_{spec}$}&
\multicolumn{1}{c}{ Log(g)}&
\multicolumn{1}{c}{ $\chi$}&
\multicolumn{1}{c}{[Fe/H]} \\
\multicolumn{1}{l}{} &
\multicolumn{1}{c}{K}&
\multicolumn{1}{c}{ cms$^{-2}$}&
\multicolumn{1}{c}{ kms$^{-1}$}&
\multicolumn{1}{c}{}
}
\startdata
102531    & 6238 $\pm$  59 & 3.80  $\pm$ 0.16 & 1.85  $\pm$ 0.05  &  0.07 $\pm$ 0.04 \\
103983    & 5750 $\pm$  56 & 4.52  $\pm$ 0.20 & 1.16  $\pm$ 0.06  &  0.02 $\pm$ 0.05 \\
105341$^a$& 4005 $\pm$  88 & 4.67  $\pm$ 0.28 & 0.83  $\pm$ 0.60  & -0.05 $\pm$ 0.19 \\
11033     & 4510 $\pm$  69 & 2.40  $\pm$ 0.30 & 1.60  $\pm$ 0.08  &  0.12 $\pm$ 0.04 \\
112222    & 6369 $\pm$ 100 & 4.10  $\pm$ 0.26 & 1.69  $\pm$ 0.08  &  0.04 $\pm$ 0.07 \\
112447    & 6095 $\pm$ 100 & 3.75  $\pm$ 0.25 & 1.82  $\pm$ 0.16  & -0.34 $\pm$ 0.08 \\
113622    & 4295 $\pm$  86 & 2.10  $\pm$ 0.25 & 1.52  $\pm$ 0.13  &  0.00 $\pm$ 0.08 \\
114155$^a$& 4348 $\pm$  75 & 1.34  $\pm$ 0.25 & 2.19  $\pm$ 0.08  & -0.58 $\pm$ 0.07 \\
\enddata
\label{basic}
\end{deluxetable}
$^a$ Surface gravities for these stars are physical, calculated as discussed
in the test.\\
$^b$ Surface gravity for this star is physical, calculated as discussed in the
text.  The microturbulence was set to 0.\\
\begin{deluxetable}{l c c c c c c c c c c c c c c c c c c c c c c c c c c c c c c c c c c c c c c c c c c c c c c c c c c c c c c c c c c c c c c c c c c c c c}
\rotate
\tablewidth{0 pt}
\tablecaption{All Equivalent Widths}
\tablehead{
\multicolumn{1}{l}{Wavelength}& 
\multicolumn{1}{c}{Ion}&
\multicolumn{1}{c}{Excitation Potential }&
\multicolumn{1}{c}{log(gf) }& 
\multicolumn{1}{c}{HIP102531  EQW    } &
\multicolumn{1}{c}{HIP102531 LogN(x) } &
\multicolumn{1}{c}{HIP105341  EQW} &
\multicolumn{1}{c}{HIP105341 LogN(x) } &
\multicolumn{1}{c}{HIP11033   EQW} &
\multicolumn{1}{c}{HIP11033  LogN(x) } &
\multicolumn{1}{c}{HIP112222  EQW} &
\multicolumn{1}{c}{HIP112222 LogN(x) } &
\multicolumn{1}{c}{HIP112447  EQW} &
\multicolumn{1}{c}{HIP112447 LogN(x) } &
\multicolumn{1}{c}{HIP113622  EQW} &
\multicolumn{1}{c}{HIP113622 LogN(x) } &
\multicolumn{1}{c}{HIP114155  EQW} &
\multicolumn{1}{c}{HIP114155 LogN(x) } &
\multicolumn{1}{c}{HIP114924  EQW} &
\multicolumn{1}{c}{HIP114924 LogN(x) } &
\multicolumn{1}{c}{HIP12784   EQW} &
\multicolumn{1}{c}{HIP12784  LogN(x) } &
\multicolumn{1}{c}{HIP13701   EQW} &
\multicolumn{1}{c}{HIP13701  LogN(x) } &
\multicolumn{1}{c}{HIP14501   EQW} &
\multicolumn{1}{c}{HIP14501  LogN(x) } &
\multicolumn{1}{c}{HIP17792   EQW} &
\multicolumn{1}{c}{HIP17792  LogN(x) } &
\multicolumn{1}{c}{HIP23852   EQW} &
\multicolumn{1}{c}{HIP23852  LogN(x) } &
\multicolumn{1}{c}{HIP29525   EQW} &
\multicolumn{1}{c}{HIP29525  LogN(x) } &
\multicolumn{1}{c}{HIP29843   EQW} &
\multicolumn{1}{c}{HIP29843  LogN(x) } &
\multicolumn{1}{c}{HIP33671   EQW} &
\multicolumn{1}{c}{HIP33671  LogN(x) } &
\multicolumn{1}{c}{HIP34440   EQW} &
\multicolumn{1}{c}{HIP34440  LogN(x) } &
\multicolumn{1}{c}{HIP3455    EQW} &
\multicolumn{1}{c}{HIP3455   LogN(x) } &
\multicolumn{1}{c}{HIP3559    EQW} &
\multicolumn{1}{c}{HIP3559   LogN(x) } &
\multicolumn{1}{c}{HIP36732   EQW} &
\multicolumn{1}{c}{HIP36732  LogN(x) } &
\multicolumn{1}{c}{HIP3992    EQW} &
\multicolumn{1}{c}{HIP3992   LogN(x) } &
\multicolumn{1}{c}{HIP40023   EQW} &
\multicolumn{1}{c}{HIP40023  LogN(x) } &
\multicolumn{1}{c}{HIP41484   EQW} &
\multicolumn{1}{c}{HIP41484  LogN(x) } &
\multicolumn{1}{c}{HIP42499   EQW} &
\multicolumn{1}{c}{HIP42499  LogN(x) } &
\multicolumn{1}{c}{HIP4346    EQW} &
\multicolumn{1}{c}{HIP4346   LogN(x) } &
\multicolumn{1}{c}{HIP43557   EQW} &
\multicolumn{1}{c}{HIP43557  LogN(x) } &
\multicolumn{1}{c}{HIP45617   EQW} &
\multicolumn{1}{c}{HIP45617  LogN(x) } &
\multicolumn{1}{c}{HIP5027    EQW} &
\multicolumn{1}{c}{HIP5027   LogN(x) } &
\multicolumn{1}{c}{HIP50505   EQW} &
\multicolumn{1}{c}{HIP50505  LogN(x) } &
\multicolumn{1}{c}{HIP5286    EQW} &
\multicolumn{1}{c}{HIP5286   LogN(x) } &
\multicolumn{1}{c}{HIP53229   EQW} &
\multicolumn{1}{c}{HIP53229  LogN(x) } &
\multicolumn{1}{c}{HIP53465   EQW} &
\multicolumn{1}{c}{HIP53465  LogN(x) } &
\multicolumn{1}{c}{HIP6732    EQW} &
\multicolumn{1}{c}{HIP6732   LogN(x) }\\
\multicolumn{1}{l}{\AA       }       &
\multicolumn{1}{c}{} &
\multicolumn{1}{c}{eV}       &
\multicolumn{1}{c}{  }&
\multicolumn{1}{c}{m\AA } &
\multicolumn{1}{c}{  }	 &
\multicolumn{1}{c}{m\AA }&
\multicolumn{1}{c}{  }	 &
\multicolumn{1}{c}{m\AA }&
\multicolumn{1}{c}{  }	 &
\multicolumn{1}{c}{m\AA }&
\multicolumn{1}{c}{  }	 &
\multicolumn{1}{c}{m\AA }&
\multicolumn{1}{c}{  }	 &
\multicolumn{1}{c}{m\AA }&
\multicolumn{1}{c}{  }	 &
\multicolumn{1}{c}{m\AA }&
\multicolumn{1}{c}{  }	 &
\multicolumn{1}{c}{m\AA }&
\multicolumn{1}{c}{  }&
\multicolumn{1}{c}{m\AA }&
\multicolumn{1}{c}{  }&
\multicolumn{1}{c}{m\AA }&
\multicolumn{1}{c}{  }&
\multicolumn{1}{c}{m\AA }&
\multicolumn{1}{c}{  }&
\multicolumn{1}{c}{m\AA }&
\multicolumn{1}{c}{  }&
\multicolumn{1}{c}{m\AA }&
\multicolumn{1}{c}{  }&
\multicolumn{1}{c}{m\AA }&
\multicolumn{1}{c}{  }&
\multicolumn{1}{c}{m\AA }&
\multicolumn{1}{c}{  }&
\multicolumn{1}{c}{m\AA }&
\multicolumn{1}{c}{  }&
\multicolumn{1}{c}{m\AA }&
\multicolumn{1}{c}{  }&
\multicolumn{1}{c}{m\AA }&
\multicolumn{1}{c}{  }&
\multicolumn{1}{c}{m\AA }&
\multicolumn{1}{c}{  }&
\multicolumn{1}{c}{m\AA }&
\multicolumn{1}{c}{  }&
\multicolumn{1}{c}{m\AA }&
\multicolumn{1}{c}{  }&
\multicolumn{1}{c}{m\AA }&
\multicolumn{1}{c}{  }&
\multicolumn{1}{c}{m\AA }&
\multicolumn{1}{c}{  }&
\multicolumn{1}{c}{m\AA }&
\multicolumn{1}{c}{  }&
\multicolumn{1}{c}{m\AA }&
\multicolumn{1}{c}{  }&
\multicolumn{1}{c}{m\AA }&
\multicolumn{1}{c}{  }&
\multicolumn{1}{c}{m\AA }&
\multicolumn{1}{c}{  }&
\multicolumn{1}{c}{m\AA }&
\multicolumn{1}{c}{  }&
\multicolumn{1}{c}{m\AA }&
\multicolumn{1}{c}{  }&
\multicolumn{1}{c}{m\AA }&
\multicolumn{1}{c}{  }&
\multicolumn{1}{c}{m\AA }&
\multicolumn{1}{c}{  }&
\multicolumn{1}{c}{m\AA }&
\multicolumn{1}{c}{  }&
\multicolumn{1}{c}{m\AA }&
\multicolumn{1}{c}{}}
\startdata                                                     
  6154.230 & \ion{Na}{1} &  2.10  &  -1.53  &  42.70   &  6.31    &  125.00  &  5.83    &  $\ldots$ &  $\ldots$ &  26.70   &   6.34   &   18.20  &   6.02   &  $\ldots$ &  $\ldots$ &  107.10  &  6.23    &  30.80   &  6.33    &  $\ldots$  & $\ldots$ &  38.00   &  5.63    &  40.20   &  6.29    &  $\ldots$ &  $\ldots$ &  27.40   &  6.07    &  33.80   &  6.14    &  37.80   &  6.44    &  18.30   &  5.99    &  75.50   &  6.25    &  $\ldots$ &  $\ldots$ &  20.00   &  5.93    &  $\ldots$ &  $\ldots$ &  78.40   &  6.28    &  57.10   &  6.29    &  37.50   &  6.29    &  $\ldots$ &  $\ldots$ &  $\ldots$ &  $\ldots$ &  34.10   &  6.21    &  79.50   &  6.19    &  92.70   &  5.85    &  32.70   &  6.10    &  131.80  &  6.52    &  $\ldots$ &  $\ldots$ &  $\ldots$ &  $\ldots$ &  $\ldots$ &  $\ldots$																																																							  \\
\enddata
\label{alleqw}
\end{deluxetable}                              
\begin{deluxetable}{l c c c c c c c c c c c} 
\rotate
\tablewidth{0 pt}
\tablecaption{Abundances}
\tablehead{
\multicolumn{1}{l}{HIP}    & 
\multicolumn{1}{c}{[Mg/H]}           &
\multicolumn{1}{c}{[Si/H]}            &
\multicolumn{1}{c}{[Ca/H]}           &
\multicolumn{1}{c}{[Ti/H]}            &
\multicolumn{1}{c}{[Ti2/H]}           &
\multicolumn{1}{c}{[Cr/H]}              &
\multicolumn{1}{c}{[Mn/H]}      &
\multicolumn{1}{c}{[Ni/H]}      &
\multicolumn{1}{c}{[Na/H]}      &
\multicolumn{1}{c}{[Al/H]}      &
\multicolumn{1}{c}{[Ba/H]}      \\
\multicolumn{1}{l}{}    & 
\multicolumn{1}{c}{}           &
\multicolumn{1}{c}{}            &
\multicolumn{1}{c}{}           &
\multicolumn{1}{c}{}            &
\multicolumn{1}{c}{}            &
\multicolumn{1}{c}{}           &
\multicolumn{1}{c}{}            &
\multicolumn{1}{c}{}           &
\multicolumn{1}{c}{}           &
\multicolumn{1}{c}{}            &
\multicolumn{1}{c}{}    }        
\startdata
102531  & 0.13   $\pm$ 0.08   & 0.16   $\pm$ 0.08 &  0.08 $\pm$ 0.06 & 0.07   $\pm$ 0.08  & 0.01   $\pm$ 0.08 &     -0.02  $\pm$ 0.09  & -0.14  $\pm$  0.05 & 0.05   $\pm$ 0.05 &  0.18   $\pm$ 0.03 &   0.07	$\pm$ 0.08     & -0.01  $\pm$ 0.13\\
103983  & -0.04  $\pm$ 0.09   & 0.02   $\pm$ 0.06 &  0.06 $\pm$ 0.08 & 0.03   $\pm$ 0.09  & -0.06  $\pm$ 0.26 &     -0.06  $\pm$ 0.11  & -0.25  $\pm$  0.17 & -0.02  $\pm$ 0.08 &  0.05   $\pm$ 0.05 &   -0.03  $\pm$ 0.06     & 0.10	$\pm$ 0.11\\
105341  & 0.00   $\pm$ 0.12   & $\ldots$          &  $\ldots$        & -0.17  $\pm$ 0.15  & $\ldots$          &     $\ldots$           & -0.20  $\pm$  0.12 & $\ldots$          &  $\ldots$          &   0.01	$\pm$ 0.19     &  $\ldots$	      \\
11033   & 0.24   $\pm$ 0.12   & 0.40   $\pm$ 0.13 &  0.13 $\pm$ 0.14 & 0.04   $\pm$ 0.17  & $\ldots$          &     -0.01  $\pm$ 0.10  & $\ldots$           & 0.23   $\pm$ 0.19 &  0.23   $\pm$ 0.18 &   0.25	$\pm$ 0.13     &  $\ldots$	      \\
112222  & 0.11   $\pm$ 0.11   & 0.04   $\pm$ 0.10 &  0.10 $\pm$ 0.08 & 0.06   $\pm$ 0.09  & -0.01  $\pm$ 0.12 &     -0.01  $\pm$ 0.13  & -0.25  $\pm$  0.09 & 0.05   $\pm$ 0.13 &  0.16   $\pm$ 0.12 &   -0.05  $\pm$ 0.09     & -0.07  $\pm$ 0.13\\
112447  & -0.21  $\pm$ 0.12   & -0.24  $\pm$ 0.17 &  0.17 $\pm$ 0.07 & -0.24  $\pm$ 0.15  & -0.41  $\pm$ 0.11 &     -0.35  $\pm$ 0.15  & -0.75  $\pm$  0.15 & -0.28  $\pm$ 0.14 &  -0.23  $\pm$ 0.11 &   0.17	$\pm$ 0.09     & -0.46  $\pm$ 0.12\\
113622  & 0.18   $\pm$ 0.13   & -0.01  $\pm$ 0.09 &  0.09 $\pm$ 0.14 & 0.04   $\pm$ 0.17  & $\ldots$          &     0.03   $\pm$ 0.19  & 0.03   $\pm$  0.20 & 0.22   $\pm$ 0.26 &  $\ldots$	     &   0.05	$\pm$ 0.22     & -0.04  $\pm$ 0.21    \\
114155  & $\ldots$            & $\ldots$          &  $\ldots$        & -0.51  $\pm$ 0.24  & $\ldots$          &     -0.61  $\pm$ 0.20  & -0.62  $\pm$  0.09 & -0.62  $\pm$ 0.14 &  -0.05  $\pm$ 0.08 &   -0.07  $\pm$ 0.11     &  $\ldots$	      \\
\enddata
\label{abundances}
\end{deluxetable}
\begin{deluxetable}{l c c}
\tablewidth{0 pt}
\tablecaption{Sample Abundance Sensitivity for [Fe/H]}
\tablehead{
\multicolumn{1}{l}{Star}    & 
\multicolumn{1}{c}{Parameter}           &
\multicolumn{1}{c}{Abundance Sensitivity} \\
\multicolumn{1}{l}{}    & 
\multicolumn{1}{c}{}           &
\multicolumn{1}{c}{ } 
}
\startdata
HIP3455 & $\Delta$T = $\pm$150             & $\mp$0.02\\
        & $\Delta$log \emph{g} = $\pm$0.12 & $\mp$0.01\\
        & $\Delta\zeta$ = $\pm$0.60        & $\pm$0.02\\
	& $\Delta<$[Fe/H]$>$               & $\pm$0.01\\
	& Total Uncertainty in [Fe/H]      & $\pm$0.03\\
\hline
HIP3559 & $\Delta$T = $\pm$150             & $\mp$0.03\\
        & $\Delta$log \emph{g} = $\pm$0.12 & $\mp$0.00\\
        & $\Delta\zeta$ = $\pm$0.60        & $\pm$0.01\\
	& $\Delta<$[Fe/H]$>$               & $\pm$0.01\\
	& Total Uncertainty in [Fe/H]      & $\pm$0.03\\
\hline
HIP103983 & $\Delta$T = $\pm$150           & $\mp$0.04\\
        & $\Delta$log \emph{g} = $\pm$0.12 & $\mp$0.02\\
        & $\Delta\zeta$ = $\pm$0.60        & $\pm$0.01\\
	& $\Delta<$[Fe/H]$>$               & $\pm$0.01\\
	& Total Uncertainty in [Fe/H]      & $\pm$0.05\\
\enddata
\label{errors}
\end{deluxetable}
\begin{deluxetable}{l c c c c c c c c c c}
\rotate
\tablewidth{0 pt}
\tablecaption{Oxygen Abundances}
\tablehead{
\multicolumn{1}{l}{HIP}& 
\multicolumn{1}{c}{EW$_{7771}$}&
\multicolumn{1}{c}{EW$_{7774}$}&
\multicolumn{1}{c}{EW$_{7775}$} &
\multicolumn{1}{c}{LTE [O/H]$_{7771}$}           &
\multicolumn{1}{c}{LTE [O/H]$_{7774}$}           &
\multicolumn{1}{c}{LTE [O/H]$_{7775}$}           &
\multicolumn{1}{c}{[O/H]$_{6300}$}&
\multicolumn{1}{c}{NLTE [O/H]$_{7771}$}           &
\multicolumn{1}{c}{NLTE [O/H]$_{7774}$}           &
\multicolumn{1}{c}{NLTE [O/H]$_{7775}$} \\
\multicolumn{1}{l}{}& 
\multicolumn{1}{c}{m{\AA}}&
\multicolumn{1}{c}{m{\AA}}&
\multicolumn{1}{c}{m{\AA}}&
\multicolumn{1}{c}{}           &
\multicolumn{1}{c}{}           &
\multicolumn{1}{c}{} 	       &	
\multicolumn{1}{c}{} 	       &	
\multicolumn{1}{c}{}           &
\multicolumn{1}{c}{}           &
\multicolumn{1}{c}{} }
\startdata
102531  & 157.7   & 139.7   & 114.0   &  0.48	$\pm$ 0.04  & 0.44   $\pm$ 0.04  & 0.33   $\pm$ 0.05 & 0.18  $\pm$ 0.07 & 0.11    $\pm$ 0.04  &  0.13	$\pm$ 0.04  &0.10   $\pm$ 0.05 \\
103983  &  67.8   &  60.8   &  48.0   &  -0.03  $\pm$ 0.09  & 0.00   $\pm$ 0.09  & 0.01   $\pm$ 0.11 & $\ldots$ 	& 0.00    $\pm$ 0.09  &  0.01	$\pm$ 0.09  &0.02   $\pm$ 0.11 \\
11033   &  23.0   &  37.8   &  27.1   &  -0.06  $\pm$ 0.14  & 0.62   $\pm$ 0.15  & 0.48   $\pm$ 0.14 & 0.19  $\pm$ 0.14 & -0.03   $\pm$ 0.14  &  0.63	$\pm$ 0.15  &0.48   $\pm$ 0.14 \\
112222  & 129.8   & 117.5   & 101.0   &  0.16	$\pm$ 0.09  & 0.18   $\pm$ 0.09  & 0.17   $\pm$ 0.14 & $\ldots$ 	& -0.05   $\pm$ 0.09  &  -0.02  $\pm$ 0.09  &0.00   $\pm$ 0.14 \\
112447  & 109.1   &  96.3   &  76.6   &  0.01	$\pm$ 0.09  & 0.01   $\pm$ 0.09  & -0.06  $\pm$ 0.11 & $\ldots$ 	& -0.02   $\pm$ 0.09  &  -0.14  $\pm$ 0.09  &-0.07  $\pm$ 0.11 \\
113622  &  19.5   &  45.4   &  26.5   &  0.15	$\pm$ 0.28  & 1.06   $\pm$ 0.28  & 0.69   $\pm$ 0.29 & 0.04  $\pm$ 0.15 & 0.18    $\pm$ 0.28  &  1.00	$\pm$ 0.28  &0.68   $\pm$ 0.29 \\
114155  &  44.2   &  52.0   &  35.3   &  0.39	$\pm$ 0.37  & 0.75   $\pm$ 0.37  & 0.53   $\pm$ 0.18 & -0.59 $\pm$ 0.11 & 0.32    $\pm$ 0.37  &  0.65	$\pm$ 0.37  &0.47   $\pm$ 0.18 \\
114924  & 109.2   &  97.6   &  78.2   &  0.11	$\pm$ 0.03  & 0.14   $\pm$ 0.03  & 0.08   $\pm$ 0.04 & $\ldots$ 	& 0.03    $\pm$ 0.03  &  0.07	$\pm$ 0.03  &0.03   $\pm$ 0.04 \\
\enddata
\label{oxygen}
\end{deluxetable}
\begin{deluxetable}{l c}
\tablewidth{0 pt}
\tablecaption{Lithium}
\tablehead{
\multicolumn{1}{l}{HIP}    & 
\multicolumn{1}{c}{LogN(Li)}          \\
\multicolumn{1}{c}{}    & 
\multicolumn{1}{c}{}          \\}
\startdata
3559    & 2.45        $\pm$ 0.03 \\
5027    & $\le$ -0.20            \\
5286    & $\le$ 0.70             \\
14501   & $\le$ 0.30             \\
23852   & 2.00        $\pm$ 0.04 \\
29525   & 2.03        $\pm$ 0.02 \\
29843   & $\le$ 1.60             \\
33671   & 2.61        $\pm$ 0.05 \\
41484   & 1.93        $\pm$ 0.04 \\
42499   & $\le$ 1.09             \\
43557   & $\le$ 1.50             \\
45617   & $\le$ 0.60             \\
50505   & $\le$ 0.50             \\  			 
102531  & 1.90        $\pm$ 0.07 \\
103983  & 1.90        $\pm$ 0.07 \\
105341  & $\le$ -0.25            \\
112447  & 2.30        $\pm$ 0.07 \\
114924  & 2.75        $\pm$ 0.06 \\
112222  & $\le$ 1.22             \\
\enddata
\label{lithium_table}
\end{deluxetable}
\begin{deluxetable}{l c c}
\tablewidth{0 pt}
\tablecaption{Membership Status}
\tablehead{
\multicolumn{1}{l}{HIP}    & 
\multicolumn{1}{c}{T$_{spec}$}     &
\multicolumn{1}{c}{ [Fe/H]}           \\
\multicolumn{1}{c}{}                      &
\multicolumn{1}{c}{K       }       &
\multicolumn{1}{c}{}
}\startdata
UNLIKELY &                &                    \\
3559     & 5800 $\pm$  38 &   -0.18 $\pm$ 0.03 \\ 
4346*    & 3820 $\pm$ 200 &    0.24 $\pm$ 0.18 \\
5027**   & 4398 $\pm$  79 &   -0.08 $\pm$ 0.17 \\
5286     & 4683 $\pm$  74 &    0.29 $\pm$ 0.07 \\
11033    & 4510 $\pm$  69 &    0.12 $\pm$ 0.04 \\
17792    & 4416 $\pm$  32 &   -0.52 $\pm$ 0.06 \\
23852    & 5778 $\pm$  38 &   -0.14 $\pm$ 0.06 \\
\enddata
\label{membership}
\end{deluxetable}
\begin{deluxetable}{l c c c c}
\tablewidth{0 pt}
\tablecaption{Group Abundances}\label{group}
\tablehead{
\multicolumn{1}{l}{Element}    & 
\multicolumn{1}{c}{[X/H]}     &
\multicolumn{1}{c}{${\chi_{\nu}^2}_{all}$}     &
\multicolumn{1}{c}{${\chi_{\nu}^2}_{group}$}     &
\multicolumn{1}{c}{${\chi_{\nu}^2}_{probable}$}\\
\multicolumn{1}{c}{}                      &
\multicolumn{1}{c}{}                      &
\multicolumn{1}{c}{}                      &
\multicolumn{1}{c}{}                      &
\multicolumn{1}{c}{}
}
\startdata
Al/H   & -0.01 $\pm$ 0.01 & 4.71      & 2.46   & 2.96 \\ 
Ba/H   &  0.00 $\pm$ 0.02 & 3.82      & 1.12   & 1.60 \\
Ca/H   &  0.09 $\pm$ 0.02 & 0.42      & 0.45   & 0.13 \\
Cr/H   & -0.10 $\pm$ 0.02 & 12.54     & 2.78   & 0.43 \\
Fe/H   & -0.01 $\pm$ 0.01 & 10.40     & 2.77   & 1.06 \\
Mg/H   &  0.04 $\pm$ 0.02 & 2.12      & 1.60   & 1.11 \\
Mn/H   & -0.11 $\pm$ 0.02 & 10.39     & 2.67   & 1.59 \\
Na/H   &  0.02 $\pm$ 0.01 & 7.46      & 4.70   & 4.97 \\
Ni/H   & -0.03 $\pm$ 0.02 & 4.09      & 1.92   & 1.31 \\
Si/H   &  0.02 $\pm$ 0.02 & 4.82      & 0.77   & 0.84 \\
Ti/H   & -0.01 $\pm$ 0.02 & 1.46      & 0.67   & 0.21 \\
Ti2/H  & -0.01 $\pm$ 0.02 & 2.72      & 0.77   & 1.28 \\
\enddata
\label{group}
\end{deluxetable}

\appendix

\section{Notes on Individual Stars: Unlikely Members}

\subsection{HIP 3559: T=5800 log{\rm }g=4.07 $\xi$=1.27 [Fe/H]=-0.18}

This star resides above the ZAMS in the HR diagram.  Ca II H and K measurements indicate an
inactive chromosphere (logR'$_{HK}$=-5.16); the activity-age calibrations of \citet{2008ApJ...687.1264M} 
suggest an age of 9.4 $\pm$ 2.7 Gyr, indicating the HR diagram position is not a result of being 
a PMS star.  Indeed, \citet{2007A&A...475..519H} derive an isochrone age, from Padova isochrones 
(\citet{2000A&AS..141..371G}, \citet{2000A&A...361.1023S}), of 7.6 Gyr; again, clearly not PMS.  
Fitting the position of this star using Yale-Yonsei isochrones \citep{2004ApJS..155..667D}, it appears to lie near the turnoff for a 
6.8 $\pm$ 0.4 Gyr isochrone.  In addition, our spectroscopic surface gravity (log $g$$\approx$4.07) is consistent with 
subgiant status.  In order to determine if the lithium abundance (logN(Li)=2.45) is consistent
with a subgiant abundance, we obtain a reasonable
estimate of the initial lithium abundance as a ZAMS dwarf and then compare the inferred lithium
dilution with theoretical calculations.  If the star is a 
6.8 Gyr subgiant, Yale-Yonsei isochrones yield a mass of 1.10 $\pm$ 0.2 M$_{\odot}$.  Assuming this mass, as a ZAMS star of
Pleiades age ($\approx$ 120 Myr), HIP 3559 would have had a main-sequence temperature
of 6158 K.  From the lithium abundance trend traced by the Pleiades (\ref{lithium}), we infer that this
star would have possessed an abundance of logN(Li)$\sim$3.00 as a 6158 K ZAMS star.  Assuming this as the ZAMS lithium, the observed
dilution of $\sim$0.55 dex is consistent with predicted lithium dilution calculations for a 1.11 M$_{\odot}$ at an age
of $\sim$ 6.8 Gyr, performed using the Clemson-American University of Beirut stellar evolution code.
HIP3559 is therefore removed from consideration as a member of a 2-3 Gyr Wolf group.
Instead, it is assumed that this is an $\sim$ 7.0 Gyr subgiant.

\subsection{HIP 4346: T=3820 log{\rm }g=1.39 $\xi$=1.33 [Fe/H]=0.24}

The metallicity of [Fe/H]=0.24 $\pm$ 0.18 is high compared to the near zero modal value of our sample and
rests outside of the abundance band for the full sample.  While the uncertainty in our [Fe/H] value is substantial, this star
is clearly metal rich across all elements.  Considering this clear metal richness across
all abundances, this star is considered an unlikely member of a dominant chemical group comprised by our sample.

\subsection{HIP 5027: T=4398 log{\rm }g=4.70 $\xi$=0.00 [Fe/H]=-0.08}
Hip 5027 has an Fe abundance which is consistent with the dominant [Fe/H] values exhibited by our sample.  However, the
abundances of other elements (Na, Al, Mn, Ni, Mg and Si) are all markedly sub-solar, and rest outside of the abundance bands 
for the full sample.  The lithium upper limit of logN(Li) $\le$ -0.20 may place the star in the trend traced by the 
Pleiades (Figure \ref{lithium}); however the significant spread in the Pleiades lithium abundances as a function of 
decreasing temperature makes a 
firm conclusion regarding age difficult to draw.  With the majority of elements disagreeing with the dominant abundance trends of 
the entire sample, this star is classified as an unlikely member of a chemically dominant group.

\subsection{HIP 5286: T=4683 log{\rm }g=4.56 $\xi$=0.54 [Fe/H]=0.29}

HIP 5286 is a member of the high metallicity ``bump'' at [Fe/H]=$\sim$0.30 in Figure \ref{fe_smooth}.
In examining the HR diagram, HIP 5286 rests above the main sequence. 
The lithium upper limit of logN(Li)$\le$0.70 places the 
star below the Pleiades trend, suggesting that it is not a young, PMS star.  
Coincidence with the other open cluster trends is uncertain as the literature lithium abundances do 
not extend to sufficiently cool temperatures, indicating the need for lithium abundance
determinations in cool stars in intermediate age open clusters.  Examining the abundances of other elements,
HIP 5286 is clearly a metal rich star with [Fe/H]=0.29 $\pm$ 0.07, well outside of the dominant
abundance bands.  This star as an unlikely member of a chemically dominant group.

\subsection{HIP 11033: T=4510 log{\rm }g=2.40 $\xi$=1.60 [Fe/H]=0.12}

The metallicity of HIP 11033 ([Fe/H] = 0.12 $\pm$ 0.04) places this star outside of the [Fe/H] band
used to constrain homogeneity in abundance.  In examining abundances for other elements, this star is seen to reside
outside of the homogeneous bands for Al, Mg and Si and it barely resides inside the band for Na.  
While the other elements are within the abundance band, it is due primarily to the significant uncertainties
associated with the respective abundances.  With a metallicity that is inconsistent with homogeneity and
considering that half of the remaining abundances are inconsistent with the sample, this is considered to
be an unlikely member of a chemically homogeneous group.  

\subsection{HIP 17792: T=4416 log{\rm }g=2.09 $\xi$=1.50 [Fe/H]=-0.52}

HIP 17792, with [Fe/H]=-0.52 $\pm$ 0.06, is one of the stars comprised by
the low metallicity bump in Figure \ref{fe_smooth}.  This low metallicity extends
across all the Fe peak elements.  The high Al abundance ([Al/Fe]=0.40) is of note in that this star shows
similar large enhancements to those seen in red giants in some open cluster (\citet{2009ApJ...701..837S}).  
The consistently low abundances of Fe, Fe peak elements and most $\alpha$ elements
lead to the conclusion that this star is an unlikely candidate that is part of a dominant chemical subsample.  

\subsection{HIP 23852: T=5778 log{\rm }g=4.22 $\xi$=1.22 [Fe/H]=-0.14}

This star resides above the ZAMS, raising the question of pre-main sequence or subgiant status.
An isochrone age of 8.8 Gyr was estimated from Padova isochrones by
\citet{2004A&A...418..989N}.  Using Yonsei-Yale isochrones, we find an age of 7.9 $\pm$ 0.10 Gyr.  
Our spectroscopic surface gravity of the star, log $g$=4.22, is consistent with a super ZAMS 
classification.  Using a stellar mass of 1.08 M$_{\odot}$, inferred from the Yale-Yonsei isochrones, 
the ZAMS temperature of this star would have been 6052 K.  The ZAMS lithium abundance, 
inferred from the Pleiades trend of Figure \ref{lithium},
logN(Li)=3.00, suggests a factor of 10 lithium depletion, consistent with theoretical
calculations.  The current lithium abundance of logN(Li)=2.00 is too low for a PMS star, and
appears consistent with the M67 Li-T$_{\rm eff}$ trend, which implies the star is an $\sim$ 5 Gyr dwarf or
mildly evolved subgiant.  The metallicity lays outside of the abundance band used for judging homogeneity of 
the sample.  The
Fe peak elements, Na, Al and Ba all reside outside of the dominant abundance bands.  
This star is, therefore, not considered a member of a dominant chemical subsample of 2-3 Gyr in age.  

\subsection{HIP 33671: T=6040 log{\rm }g=4.40 $\xi$=1.38 [Fe/H]=-0.21}

The metallicity of HIP 33671 is [Fe/H]=-0.21 $\pm$ 0.04.  This places the star well outside 
of the apparent dominant metallicity band.  The metal poor nature applies across all 
other elements, with the star not resting within any abundance bands.  While its Li abundance is
not inconsistent with 2-3 Gyr age (Figure \ref{lithium}),
HIP 33671 is unlikely to be part of a dominant chemical subsample.

\subsection{HIP 42499: T=4994 log{\rm }g=4.41 $\xi$=0.59 [Fe/H]=-0.56}

HIP 42499 is a member of the metal-weak peak in the full sample 
[Fe/H] distribution (Figure \ref{fe_smooth}).  A Li upper limit of
logN(Li)$\le$1.09 potentially places this star in the Hyades trend, however
the chromospheric activity (logR'$_{HK}$=-4.98) is much lower than the activity trend
for the Hyades, suggesting it is older than the Hyades.  
With an [Fe/H]=-0.56 $\pm$ 0.10 and consistent metal deficiency evinced across all 
elements, this star is classified as 
an unlikely member of a chemically dominant group.

\subsection{HIP45617 T=4855 log{\rm }g=4.35 $\xi$=1.01 [Fe/H]=-0.12}

This star resides above the lower main sequence. 
The lithium upper limit (logN(Li) $\le$ 0.60) shows that the star 
is not a pre-main sequence object.  The activity of logR'$_{HK}$=-4.60, from the Ca II H and K
survey of the solar neighborhood of D. Soderblom (private communication), would place the star
below the activity trend of the Hyades, qualitatively suggesting that a Hyades age would be
a reasonable lower limit.  However, the spectroscopic surface gravity is somewhat low for a dwarf star.
A possible explanation is that overionization, observed in many cool cluster dwarfs (\citet{2003AJ....125.2085S}),
is yielding spuriously low surface gravities.  With a greater number of atoms in ionized states, the gravity would have
to be artificially lowered to obtain ionization balance.  However
excellent agreement is seen between the spectroscopic gravity (log $g$=4.35) and the physical
gravity (log $g$=4.38).  The star's [Fe/H]=-0.12 $\pm$ 0.02 is inconsistent with it being
a member of the dominant metallicity distribution and, it does not
reside within the abundance bands for any other elements.  While its low surface gravity remains
a mystery, we consider HIP 45617 an unlikely member of a dominant chemical group.

\subsection{HIP 50505: T=5655 log{\rm }g=4.42 $\xi$=1.16 [Fe/H]=-0.14}

This star clearly resides on the main sequence, with a low Li upper limit of 
logN(Li)$\le$0.50, consistent with the star being an old ($\le$ M67 age) dwarf. 
The star is clearly metal poor ([Fe/H]=-0.14 $\pm$ 0.03) when 
compared with our sample mean metallicity.  In examining all other elements, 
the star rests outside of the group abundance bands for Fe, the Fe peak elements and for Si.  
The tightly constrained metallicity and the consistently low abundances across all Fe 
peak elements lead to classification of this star as an unlikely group member of a dominant metallicity group in our sample.  

\subsection{HIP 112447: T=6095 log{\rm }g=3.75 $\xi$=1.82 [Fe/H]=-0.34}

This star has a distinctly low [Fe/H]=-0.34 $\pm$ 0.08, making it a member
of the metal poor peak of Figure \ref{fe_smooth}.  The abundances of other 
elements are similarly metal poor.  With this strong evidence
for metal poverty, HIP 112447 is classified as an unlikely member of a chemically
dominant group.

\subsection{HIP 114155: T=4348 log{\rm }g=1.34 $\xi$=2.19 [Fe/H]=-0.58}

The [Fe/H] of HIP 114155 is clearly low [Fe/H]=$-$0.58 $\pm$ 0.07.  This giant
shows similarly low abundances of all elements with the exception of [Na/H] and [Al/H].  
The enhanced Na and Al abundance ratios ([Na/Fe]=0.53 $\pm$ 0.08 and [Al/Fe]=0.51 $\pm$ 0.11) can 
be compared with those in open cluster giants.  In a recent analysis of Hyades dwarfs
and giants, \citet{2009ApJ...701..837S} found enhancements in Na and Al of between 0.20 and 0.50 dex in cluster
giants compared with dwarfs, a result in conflict with standard stellar models.  Similar abundance
enhancements are seen in other open clusters \citep{2007AJ....134.1216J}.  This points
to a pattern of anomalously large Na and Al abundances in population I giants, likely
a side effect of NLTE effects, discussed in more detail below.  Regardless, the distinctly
low metal abundances across multiple elements suggest that HIP114155 is an unlikely member
of the dominant chemical group in our sample.

\section{Notes on Individual Stars: Possible Members}

\subsection{HIP 3992: T=4772 log{\rm }g=2.58 $\xi$=1.59 [Fe/H]=-0.15}

HIP 3992 has an [Fe/H]=$-$0.15 $\pm$ 0.04.  This places it outside of the Fe band that appears to dominate the
sample.  However, all other abundances rest inside their respective bands, suggesting the star is
chemically consistent with the overall dominant chemical composition of our sample.  
We thus consider HIP 3992 a possible member of a dominant chemical group in our sample.

\subsection{HIP 12784: T=4701 log{\rm }g=2.68 $\xi$=1.49 [Fe/H]=0.09}

The uncertainty associated with the metallicity of HIP 12784 ([Fe/H]=0.09 $\pm$ 0.13) places it within the 
dominant Fe band.  The star resides outside of the abundance bands for Mn, Mg, Al, Na and Ni.  Thus we consider
this star only a possible member of a dominant chemical group.

\subsection{HIP 29843: T=6130 log{\rm }g=4.11 $\xi$=1.52 [Fe/H]=0.12}

HIP 29843 has a Li upper limit of logN(Li)$\le$ 1.60.  
We estimate a Yale-Yonsei isochrone mass of 1.43 $\pm$ 0.02
M$_{\odot}$.  Using this mass to determine the ZAMS temperature of 
the star yields T$_{ZAMS}$=6678 K.  This temperature, when compared to 
Figure \ref{lithium}, would have placed this star in or on the blue-edge of the 
lithium dip while a dwarf.  Currently, as a subgiant that has emerged from the lithium
dip, the lack of lithium suggests that the deepening convection zone in the subgiant has 
not brought lithium back to the surface.  This appears consistent with the
findings of \citet{1990ASPC....9..357B} who also inferred little transport of lithium to the surface
in subgiants emerging from the lithium dip in M67.  The metallicity 
of the star ([Fe/H]=0.12 $\pm$ 0.08) appears to be somewhat high compared to the peak 
of the sample, and, indeed it rests outside of the [Fe/H] 
band.  This star also resides outside of the abundance bands for Si, Na, Cr and Mn.  
However, it rests within the abundance bands for the other seven elements.
This leads us to consider HIP 29843 to be a possible member of a dominant metallicity sample.

\subsection{HIP 34440: T=4757 log{\rm }g=2.43 $\xi$=1.46 [Fe/H]=-0.15}

Fe, Ti and Cr for this star lay outside of the respective abundance bands.  The other elements all reside within
their bands, consistent with homogeneity.  We consider this star only a possible member of a dominant homogeneous
chemical group in our sample.  

\subsection{HIP 36732: T=4667 log{\rm }g=2.54 $\xi$=1.44 [Fe/H]=0.10}

Fe, Mn, Ni, Na and Mg all appear slightly enriched when compared to the dominant abundance bands.  While the other elements
have abundances within their respective bands the consistent overabundances for Fe, Mn, Ni, Na and Mg suggest this
star be classified as only a possible member of a chemically dominant subsample. 

\subsection{HIP 41484: T=5855 log{\rm }g=4.41 $\xi$=1.17 [Fe/H]=0.08}

The Fe abundance of HIP 41484 ([Fe/H]=0.08 $\pm$ 0.03) is supersolar compared to our sample mean.  This
supersolar value, however, is not consistent across the other elements.  The abundances derived for all other
elements agree with the abundance bands used to constrain homogeneity.  The lithium abundance of logN(Li)=1.93 $\pm$
0.04 places the star below the lithium trend of the Pleiades and suggests a lower age limit
of approximately Hyades age, and perhaps at least as large as the age of NGC 752 and M67.  Considering the homogeneity across multiple
elements, but not for Fe, this star is considered a possible member of a chmically homogeneous dominant 2-3 Gyr subsample.

\subsection{HIP 43557: T=5816 log{\rm }g=4.52 $\xi$=1.15 [Fe/H]=-0.03}

The [Fe/H] of HIP 43557 matches the mean abundance of our entire sample.  Mg, Na and Si however, 
do not appear to lay within their respective abundance bands.  The average abundances for Ti, \ion{Ti}{2}, Cr and
Ba all rest near the sample mean abundances irrespective of their uncertainties, suggesting a high degree of homogeneity.
In examining the lithium abundance, the star rests below the 5 Gyr trend in the Li-T$_{eff}$ relation, perhaps suggesting
an older age.
Although the [Fe/H] agrees
well with the mean metallicity and the average abundances of multiple elements are close to the respective mean
abundances for the group, the evidence from Mg, Na and Si and the lower lithium abundance make HIP 43557 a possible group member.

\subsection{HIP103983: T=5750 log{\rm }g=4.52 $\xi$=1.16 [Fe/H]=0.02}

The status of this star is somewhat of an enigma.  While an isochrone fit is consistent with
placement on the subgiant branch of an 8.5 $\pm$ 0.11 Gyr isochrone, the surface gravity of log $g$=4.52 
suggests a dwarf luminosity
class.  Note the significant uncertainty in the surface gravity measurement (0.20
dex).  \citet{2005ApJS..159..141V} find a surface gravity 
of 4.37, consistent with the lower limit of the spectroscopic gravity derived here.  Further comparing surface
gravity estimates, the physical surface gravity derived for this star is log $g$=4.22 $\pm$ 0.23, which
would be consistent with subgiant status.  The lithium
abundance, logN(Li)=1.90 $\pm$ 0.07 places this star within the lower end of the lithium
trend observed in both NGC 752 and M67.  This would suggest consistency with a 2-3 Gyr
age for a dwarf, but definitive age conclusions based on the lithium are impractical considering
the scatter and overlap in lithium abundances in both NGC 752 and M67.  If the star were indeed a subgiant,
does the lithium abundance yield different conclusions?  The mass of an 8.5 Gyr subgiant with a 
temperature of 5750 K would be 1.05 M$_{\odot}$.  This yields a ZAMS temperature, T$_{ZAMS}$=5754 K, which coincides
with a Pleiades lithium abundance of logN(Li)=3.00 on the so-called ``lithium plateau''.  Assuming this as a reasonable
ZAMS lithium abundance, this star would have $\approx$1.10 dex depletion, which is not entirely consistent
with predictions of 0.20-0.90 dex of lithium dilution for a 1 M$_{\odot}$ obtained from the Clemson-American
University of Beirut Stellar Evolution Code.  This perhaps points to the star not being a clear subgiant, however, 
the evolutionary status of this star remains uncertain.
Examining the abundances, the star has an [Fe/H]=0.02 $\pm$ 0.05, which is consistent with it 
being a member of a chemically dominant
subgroup with a characteristic metallicity near -0.03.  The $\alpha$ and Fe peak elements, likewise, 
yield abundances that reside within the respective abundance bands that are used to characterize 
homogeneity.  Considering the uncertainties in the surface gravities and the potential that the
lithium abundance negates a subgiant classification and that the abundances are homogeneous with the
rest of the sample, we consider this star a possible member of a chemically homogeneous sub-group.

\subsubsection{HIP 105341: T=4005 log{\rm }g=4.67 $\xi$=0.83 [Fe/H]=-0.05}

This star is the coolest dwarf in the sample.  The chromospheric activity 
(logR'$_{HK}$=-4.552) from \citet{2006AJ....132..161G} suggests this is a relatively active star, 
which may be consistent with PMS status, although it is not inconsistent with a main sequence age.   
The activity derived age, using the updated age-activity relation of \citet{2008ApJ...687.1264M}
is 0.85 Gyr $\pm$ 0.25 Gyr, which places the star on the main sequence, in agreement with the surface
gravity.  While this age estimate suggests the star
does not belong in a 2-3 Gyr Wolf group, the quoted error only includes
uncertainty based in the calibration relationship.  Furthermore, activity based ages, while useful in
a statistically significant sample size, may not be robust enough to constrain individual field
star ages well enough to eliminate pre-main sequence status for this star, although the surface gravity
may suggest this is not a pre-main sequence object.  The lithium upper limit 
(logN(Li)$\le$-0.25) may plausibly place the star in the lithium trend traced by the Pleiades in Figure \ref{lithium},
but without lithium abundances for more cool Pleids the picture is unclear.  
The Fe abundance of the star ([Fe/H]=-0.05 $\pm$ 0.19) 
is consistent with membership in a dominant chemical group centered on [Fe/H]=-0.03.  It also resides 
within the abundance bands for all elements with measurable abundances (Mg, Ti and Al).
Although the chemical homogeneity of this star is constrained by all available abundances,
this star is considered only a possible member of a chemically dominant 2-3 Gyr group due to the lack of abundances
across all elements.  

\subsection{HIP 114924: T=6179 log{\rm }g=4.36 $\xi$=1.59 [Fe/H]=0.06}

With [Fe/H]=0.06 $\pm$ 0.03, HIP114924 resides outside of the dominant Fe band.  It
also appears to reside outside of the bands for Cr and Na.  However, it resides
inside the bands for all other elements.  The lithium abundance, logN(Li)=2.75 $\pm$ 0.06 places 
the star along the lower envelope of the so-called lithium plateau in Figure \ref{lithium}.  This would suggest that 
HIP114924 is a good preserver of lithium.  It is below the Hyades plateau and falls on the NGC752 trend, therefore a
2-3 Gyr age is quite consistent with the Li.  
Given that Fe, Cr and Na are not consistent with our samples modal values, but that
other elements are, this star is considered only 
a possible member of the chemically dominant group.

\section{Note on Individual Stars: Likely Members}

\subsection{HIP 3455: T=4860 log{\rm }g=2.53 $\xi$=1.49 [Fe/H]=0.00}

This star has an [Fe/H]=0.00 $\pm$ 0.03 that is consistent with the dominant group metallicity.  In examining the other
elements, it resides within every abundance band, suggesting its classification as a likely member of a
homogeneous subsample.

\subsection{HIP 6732: T=4665 log{\rm }g=2.45 $\xi$=1.58 [Fe/H]=-0.03}

This star has a metallicity ([Fe/H]=-0.03 $\pm$ 0.04) which matches closely with the weighted average of our sample.  
Examining the other elements, the abundances all appear to reside within the respective group metallicity bands.  
The homogeneity demonstrated across
all elements and the agreement of [Fe/H] with the mean group abundance suggest that HIP 6732 is a likely member of
a chemically dominant subsample.

\subsection{HIP 13701: T=4675 log{\rm }g=2.71 $\xi$=1.37 [Fe/H]=-0.03}

This star clearly resides within the dominant [Fe/H] band.  Indeed, its abundance is nearly identical to the 
weighted mean of the sample.  It appears consistent with the metallicity bands for all elements.  This
homogeneity with the rest of the sample leads to classifying HIP 13701 as a likely group member.

\subsection{HIP 14501: T=5785 log{\rm }g=4.44 $\xi$=1.24 [Fe/H]=-0.08}

Having [Fe/H]=-0.08 $\pm$ 0.04, HIP 14501 resides inside of the dominant metallicity band.  In fact, it
resides in the metallicity bands for all elements and, in many cases, the average abundance of each element
nearly matches with the weighted mean used to characterize the abundance trend of the sample.  The upper limit
lithium abundance (logN(Li) $\le$ 0.30), however, perhaps suggests an age of much greater than 2-3 Gyr. 
We note this inconsistency, but based on the chemical abundances of all other elements, this star is
considered a likely member.

\subsection{HIP 29525: T=5710 log{\rm }g=4.57 $\xi$=1.28 [Fe/H]=-0.03}

This star resides on the main sequence of the isochrones in Figure \ref{HR_spec_final}.
The lithium abundance (logN(Li)=2.03 $\pm$ 0.02) places the star within the abundance trends traced by  
NGC 752 and M67, perhaps consistent with membership in a 2-3 Gyr group.  The metallicity ([Fe/H]=-0.03 $\pm$ 0.03) firmly 
places this star within the abundance band that 
dominates the sample.  The abundances of Na and Al are outside of their respective abundance bands 
but all other elements are within the bands.  We consider this a candidate for likely membership in a chemically
homogeneous and dominant subsample of 2-3 Gyr age.  

\subsection{HIP 40023: T=5290 log{\rm }g=3.77 $\xi$=1.21 [Fe/H]=-0.05}

HIP 40023 has an [Fe/H]=-0.05, which is within the metallicity band of the sample.  Indeed, its abundances across multiple
elements fit inside the respective metallicity bands.  The small spread in abundances for the star
itself and relative to the overall sample abundance bands, lead to classification of this star
as a likely member of a dominant chemically homogeneous 2-3 Gyr subsample.     

\subsection{HIP 53229: T=4690 log{\rm }g=2.61 $\xi$=1.47 [Fe/H]=-0.10}

The Fe abundance of HIP 53229 ([Fe/H]=-0.10 $\pm$0.15) is consistent with the mean abundance band of our sample.  
Examining the other abundances, Cr is the only element that does not appear
within the abundance band for the group sample.  Homogeneity is observed across all the other elements,
therefore this star is likely a member of a chemically homogeneous subgroup.  

\subsection{HIP 53465: T=4570 log{\rm }g=2.50 $\xi$=1.30 [Fe/H]=-0.08}

The metallicity of HIP 53465 ([Fe/H]=-0.08 $\pm$ 0.07) is consistent with this star being a member of the dominant chemical
subgroup.  While abundances of Al and \ion{Ti}{2} are found to lay outside of the sample abundance
bands, the remaining elements show a high degree of homogeneity.  For most elements, the abundances lay
within the abundance band.  Thus, this star is considered a likely
member of a chemically dominant group in our sample.

\subsection{HIP 102531: T=6238 log{\rm }g=3.80 $\xi$=1.85 [Fe/H]=0.07}

The metallicity of HIP 102531 ([Fe/H]=0.07 $\pm$ 0.04) is barely outside of the 3-$\sigma$ cutoff of the mean Fe 
abundance of the whole sample.  However, this star resides within the mean abundance bands of 
Al, Ba, Ca, Mg, Mn, Ni, Si, Ti and Ti 2.  In Figure \ref{lithium}, this is the warmest sample star
that has lithium, and can be seen to lay significantly beneath any trend
traced by any of the plotted open cluster dwarf abundances.  From the HR diagram, this star lies
along the early subgiant branch of a 2.7 Gyr isochrone, which indicates a mass of 1.5 $\pm$ 0.1 M$_{\odot}$.  
In comparing this star with Figure 11 of \citet{1995ApJ...446..203B}, who plot lithium abundances for open clusters
versus stellar mass, the lithium abundance for the 
derived mass appears to be between the trends for M67 and NGC 752, consistent with the estimated isochrone age
of 2.7 Gyr.  This would suggest that the star has suffered subgiant and/or main sequence Lithium dip depletion.
Recognizing that the majority of elements suggest this star is part
of a chemically homogeneous subsample, and the $\sim$ 3 Gyr age implied by isochrones and Li,
it is classified as a likely member of a dominant subsample.

\subsection{HIP 112222: T=6369 log{\rm }g=4.10 $\xi$=1.69 [Fe/H]=0.04}

HIP 112222, with [Fe/H]=0.04 $\pm$ 0.07, rests within the
dominant Fe band and the abundance bands for every other element with the exception of Mn.  
HIP112222 is located at the turnoff of a 2.7 Gyr 
isochrone in Figure \ref{HR_spec_final}.  The position along this isochrone implies a mass of $\sim$
1.3 M$_{\odot}$ consistent with this possibly being a lithium dip star, providing an 
explanation for the apparently low upper limit lithium abundance of logN(Li) $\le$ 1.22.  
Its  placement in homogeneous
abundance bands across multiple elements and apparent 2-3 Gyr isochrone age
lead to this star being considered a likely member of the dominant
subsample.

\subsection{HIP 113622: T=4295 log{\rm }g=2.10 $\xi$=1.52 [Fe/H]=0.00}

With [Fe/H]=0.00 $\pm$ 0.08, this star rests comfortably inside the dominant Fe band.  
The Ni abundance of [Ni/H]=0.29 is uncharacteristically high for our sample, but the uncertainty of 0.26 dex
is signficant, which can bring the star into close agreement with the Ni band.  Furthermore,
HIP 113622 is consistently within the metallicity bands for all other elements.  Consequently, it is classified as
a likely member of a dominant subsample.

\begin{figure}
\plotone{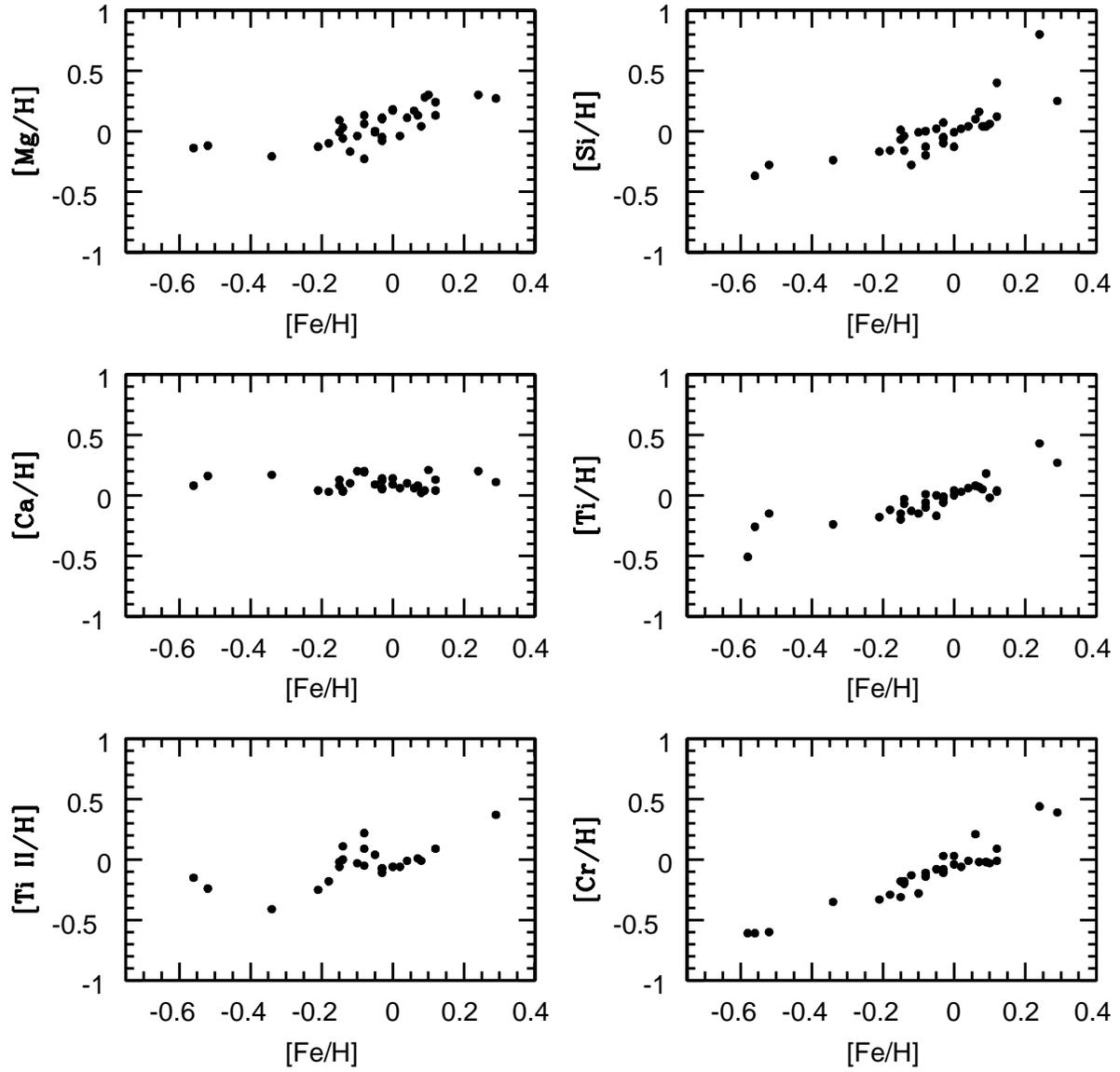}
\caption[X/H versus Fe/H]{Abundance trends for all stars versus Fe/H.}\label{abun1}
\end{figure}

\begin{figure}
\plotone{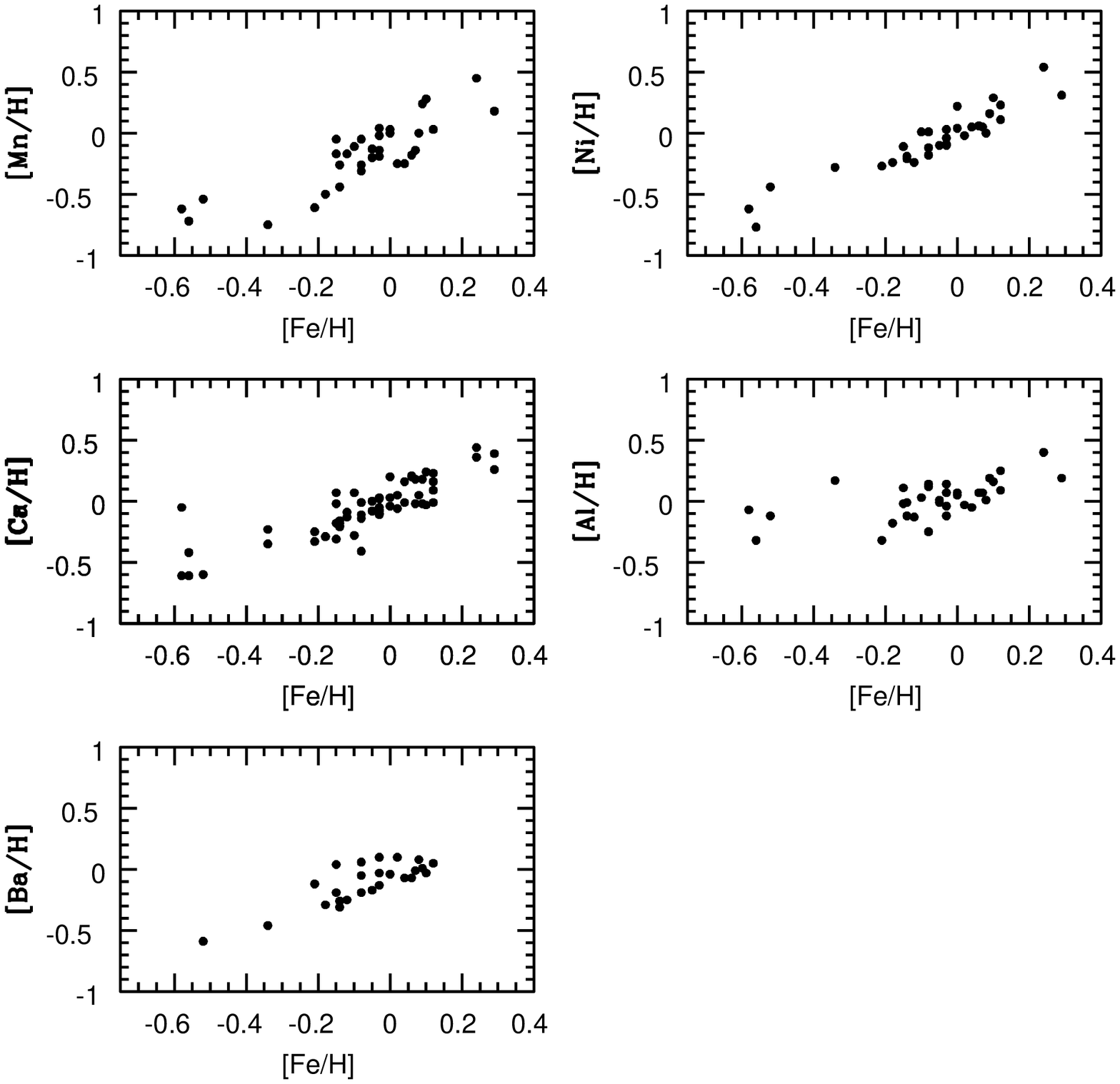}
\caption[X/H versus Fe/H]{More abundance trends for all stars versus Fe/H.}\label{abun2}
\end{figure}

\end{document}